\documentclass[seceqn,11pt]{elsart}

\usepackage{amsmath,amssymb,latexsym,amssymb,tables,psfig,mathrsfs}
\usepackage{macros,subfigure,psfrag,color,bbm}
\usepackage[font=scriptsize,labelfont=bf]{caption}
\usepackage[dvips]{epsfig}
\usepackage[dvips]{graphicx}
\usepackage{longtable,rotating}

\input eqalign.sty

\begin{document}

\begin{frontmatter}

\begin{flushright}
DESY 12-136  \\
HU-EP-12/23 \\
JLAB-THY-12-1605 \\
SFB/CPP-12-54 \\
\end{flushright}

\title{A quenched study of the Schr\"odinger functional 
with chirally rotated boundary conditions: non-perturbative tuning}

\author[HU,DESY]{J. Gonz\'alez L\'opez},
\author[DESY]{K. Jansen},
\author[JLAB]{D.\ B.\ Renner},
\author[HU]{A. Shindler}\footnote{Heisenberg Fellow}
\address[HU]{Humboldt-Universit\"at zu Berlin, Institut f\"ur Physik
  Newtonstrasse 15, 12489 Berlin, Germany}
\address[DESY]{DESY,Platanenallee 6, 15738 Zeuthen, Germany}
\address[JLAB]{Jefferson Lab, 12000 Jefferson Avenue, Newport News, VA 23606, USA}

\maketitle
\begin{abstract}
The use of chirally rotated boundary conditions provides a formulation of the Schr\"odinger functional 
that is compatible with automatic O($a$) improvement of Wilson fermions up to O($a$) boundary contributions.
The elimination of bulk O($a$) effects requires the non-perturbative tuning of the critical mass 
and one additional boundary counterterm. We present the results of such a tuning in a quenched 
setup for several values of the renormalized gauge coupling, from perturbative to non-perturbative regimes,
and for a range of lattice spacings. We also check that the correct boundary conditions and symmetries
are restored in the continuum limit. 

\end{abstract}


\end{frontmatter}
\cleardoublepage


\section{Introduction}
\label{sec:intro}

The Schr\"odinger functional of QCD~\cite{Luscher:1992an,Sint:1993un,Sint:1995rb}
is the gauge invariant functional integral for QCD on a hyper-cylinder where the fields satisfy
periodic boundary conditions in the spatial directions and Dirichlet
boundary conditions at the Euclidean times $0$ and $T$.
The SF is non-perturbatively defined and
it has been shown to be very successful when used as a
renormalization scheme for lattice QCD. An incomplete set of results obtained using the SF is given by 
refs.~\cite{Luscher:1993gh,Capitani:1998mq,Guagnelli:2003hw,Pena:2004gb,DellaMorte:2004bc,DellaMorte:2005kg,Aoki:2009tf,Aoki:2010wm,Tekin:2010mm}.

Because of the Dirichlet boundary conditions in the time direction,
the SF has a spectral gap even at zero quark mass~\cite{Sint:1993un,Luscher:2006df}, thus allowing
to use the SF as a massless renormalization scheme.
Moreover, due to the possibility of applying finite-size techniques,
the SF is an ideal framework to evaluate scale-dependent quantities over
a wide range of energies, covering the perturbative up
to the non-perturbative regimes. Such a framework is needed when studying
non-perturbative renormalization on the lattice.
Another good property of the SF is the ability to define gauge invariant quark sources,
making it possible to construct gauge invariant correlation functions to 
determine renormalization factors.

The implementation of the SF on the lattice beyond the pure gauge
theory is not a straightforward issue. Depending on the bulk lattice Dirac operator adopted,
boundary terms in the lattice action may need to be added in order to recover the 
correct boundary conditions in the continuum limit.
For the implementation with Wilson fermions~\cite{Sint:1993un,Sint:1995rb}, 
the boundary conditions arise naturally~\cite{Luscher:2006df} and no fine-tuning is needed near the boundary.
For Ginsparg-Wilson fermions, specific boundary terms have to be added to the lattice action
in order to recover the correct continuum limit~\cite{Luscher:2006df}.
A related issue is the definition of boundary conditions
in order to achieve automatic O($a$) improvement~\cite{Frezzotti:2003ni} with massless Wilson quarks.
In this case one would like to have 
boundary conditions in the continuum limit that allow Wilson fermions to maintain
automatic O($a$) improvement, similarly to what happens with twisted mass fermions. 
A solution to this problem has been proposed by Sint~\cite{Sint:2005qz,Sint:2010eh},
called the chirally rotated Schr\"odinger functional ($\chi$SF).
Being compatible with automatic O($a$) improvement, makes the $\chi$SF an ideal setup for renormalizing
bare operators computed with Wilson twisted mass fermions at maximal twist~\cite{Boucaud:2007uk}
and it may have several advantages compared with the standard SF. Here we just mention
the possibility of computing renormalization factors of four-fermion operators (like for $B_K$) or of
twist-2 operators (like for $\langle x \rangle$) free of O($a$) corrections. 

The main content of this paper is the numerical investigation of the non-perturbative 
tuning of the $\chi$SF in a quenched setup. 
In sect.~\ref{sec:Wtm} we start with Wilson twisted mass fermions as an example of automatic
O($a$) improvement. Then in sect.~\ref{sec:chiSF} we discuss the $\chi$SF 
in the continuum and its relationship with automatic O($a$) improvement. 
In sect.~\ref{sec:Wilson} we introduce the appropriate lattice action for Wilson fermions and the
boundary counterterms and in sect.~\ref{sec:tuning} we discuss the non-perturbative tuning 
of the critical mass and of the relevant boundary counterterm needed to obtain the correct continuum limit. 
In sect.~\ref{sec:scaling} we check that the proper symmetries and boundary conditions 
are recovered in the continuum limit.
In a forthcoming paper~\cite{Lopez:2012mc} we will study the application of the $\chi$SF 
renormalization scheme to the determination of several physically relevant quantities.
The computation of such quantities will moreover allow us to perform
a continuum limit scaling test for cutoff effects both for small and large volume calculations.

\section{Wilson twisted mass fermions and automatic O($a$) improvement}
\label{sec:Wtm}

The Wilson twisted mass (Wtm) formulation~\cite{Frezzotti:2000nk,Shindler:2007vp} 
is a lattice action that provides a solid framework
to perform large scale simulations with $N_f=0,2$~\cite{Jansen:2003ir,Jansen:2005kk,Boucaud:2007uk,Boucaud:2008xu} 
and more recently $N_f=2+1+1$~\cite{Baron:2010bv} flavors of dynamical fermions.
One of its advantages is the automatic O($a$) improvement of physical correlation
functions~\cite{Frezzotti:2003ni}, which requires the non-perturbative tuning of just a single parameter: 
the critical mass $m_{\rm cr}$.
The resulting lattice action is referred to as Wtm at maximal twist. 
Clearly it would be desirable to retain automatic O($a$) improvement thorugh the process of renormalizing local operators.

We now briefly summarize how automatic O($a$) improvement works for Wilson twisted mass fermions in a finite volume 
without boundaries. The very same mechanism will be used in the next section to show how the $\chi$SF
retains this property.
The action for twisted mass QCD (tmQCD) in the continuum for a flavor doublet $\chi$ of fermions is 
\be
S = \int\!\!d^4x~\overline{\chi}(x)\left[\gamma_\mu \D_\mu + m_{\rm q} + i\mu_{\rm q}\gamma_5 \tau^3\right]\chi(x),
\label{eq:tmQCD}
\ee
where $m_{\rm q}$ is the so-called untwisted mass, $\mu_{\rm q}$ the twisted mass and $\tau^3$
is the third of the Pauli matrices $\tau^{a}$. Performing the following non-anomalous change of basis
\be
\psi(x) = {\rm e}^{i\omega\gamma_5\tau^3/2}\chi(x), \qquad 
\overline{\psi}(x) = \overline{\chi}(x){\rm e}^{i\omega\gamma_5\tau^3/2},  \qquad 
\omega = \arctan\left(\frac{\mu_{\rm q}}{m_{\rm q}}\right)\,,
\label{eq:twist}
\ee
it is easy to see that the tmQCD action~\eqref{eq:tmQCD} is equivalent to the standard QCD action for degenerate
$N_f=2$ fermions $\psi$ with mass $M=\sqrt{m_{\rm q}^2 + \mu_{\rm q}^2}$.
This trivial change of basis becomes non-trivial
once we decide to discretize the QCD action with Wilson fermions obtaining
\be
S = a^4\sum_x \overline{\chi}(x)\left[\D_W + m_0 + i\mu_{\rm q}\gamma_5 \tau^3\right]\chi(x)\,,
\label{eq:Wtm}
\ee
where $\D_W$ is the standard Wilson operator
\be
\D_W = \frac{1}{2}\left[\left(\nabla_\mu + \nabla_\mu^* \right)\gamma_\mu - a \nabla_\mu^*\nabla_\mu\right]\,.
\label{eq:wilson}
\ee
The Wtm action~\eqref{eq:Wtm} has the proper continuum limit~\cite{Frezzotti:2000nk} 
and, after tuning the bare untwisted mass $m_0$ to its critical value $\mcr$, 
physical correlation functions are automatically O($a$) improved~\cite{Frezzotti:2003ni}\footnote{For other 
proofs of automatic O($a$) improvement 
see refs.~\cite{Aoki:2004ta,Sharpe:2004ny,Frezzotti:2005gi,Shindler:2005vj,Sint:2005qz,Aoki:2006nv}.}.
The property of automatic O($a$) improvement of physical correlation functions is a consequence of the 
different transformation properties of the mass term and the Wilson term under vector and axial symmetries.

We start by noting that the Wtm lattice action~\eqref{eq:Wtm} is invariant under the transformation
$\mathcal{R}^{1,2}_5 \times \widetilde{\mathcal{D}}^{1,2}$ where
\be
\mathcal{R}^{1,2}_5 \colon
\begin{cases}
\chi(x) \rightarrow i \gamma_5 \tau^{1,2} \chi(x) \\
\chibar(x) \rightarrow  \chibar(x) i \gamma_5 \tau^{1,2}\,
\end{cases}
\label{eq:R512}
\ee
is a discrete chiral transformation and
\be
\widetilde{\mathcal{D}}^{1,2} \colon
\begin{cases}
U(x;\mu) \rightarrow U^\dagger(-x-a\hat{\mu};\mu) \\ 
\chi(x) \rightarrow i\tau^{1,2}{\rm e}^{3 i \pi/2} \chi(-x) \\
\chibar(x) \rightarrow  \chibar(-x)\left(-i\tau^{1,2}\right){\rm e}^{3 i \pi/2}\,,
\end{cases}
\ee
is a discrete vector transformation combined with a transformation that essentially 
counts the dimensions of the fields~\cite{Frezzotti:2003ni}.
We consider a general multiplicatively renormalizable multilocal lattice 
field $\Phi$ that is even under the transformation $\mathcal{R}^{1,2}_5 \times \widetilde{\mathcal{D}}^{1,2}$.
In the following we will refer to even (odd) operators and correlation functions under a transformation 
if they are invariant (change sign) under that transformation.
The discretization errors of the lattice correlation function $\langle \Phi \rangle$ are described by the 
Symanzik effective theory~\cite{Symanzik:1983dc}.
The Symanzik effective action corresponding to~\eqref{eq:Wtm} (with $m_0=m_{\rm cr}$) reads
\be
S_{\rm eff} = S_0 + aS_1 + \ldots\,,
\ee
where the target continuum theory is
\be
S_0 = \int\!\!d^4x~\chibar(x) \left[ \gamma_\mu D_\mu + i\mu_{\rm q}\gamma_5 \tau^3 \right] \chi(x)\,.
\label{eq:tmQCD}
\ee
We recall that the effective theory is constructed taking into account the symmetries of the lattice 
action~\cite{Luscher:1996sc}. This implies that the higher-dimensional correction terms
in the effective action
\be
S_1 = \int\!\!d^4y~{\mathcal L}_1(y) \qquad {\mathcal L}_1(y) = \sum_i c_i
  {\mathcal O}_i(y)\,
\ee
are $\mathcal{R}^{1,2}_5 \times \widetilde{\mathcal{D}}^{1,2}$ even. 
After using the equations of motion, the only operators ${\mathcal O}_i$ contributing to on-shell correlation
 functions for vanishing untwisted quark mass are~\cite{Luscher:1996sc,Frezzotti:2001ea}
\be
  i \chibar\sigma_{\mu\nu}F_{\mu\nu}\chi,\qquad 
  \mu_{\rm q}^2\chibar \chi\,,
\label{eq:sym_op}
\ee
where $\sigma_{\mu\nu} = \frac{i}{2}\left[\gamma_\mu,\gamma_\nu\right]$ and $F_{\mu\nu}$ is the gluon field strength tensor.
We observe that, even if both $S_0$ and $S_1$ are invariant under 
$\mathcal{R}^{1,2}_5 \times \widetilde{\mathcal{D}}^{1,2}$, the continuum theory is separately invariant
under $\mathcal{R}^{1,2}_5$ and $\widetilde{\mathcal{D}}^{1,2}$,
while ${\mathcal L}_1$ is odd under both $\mathcal{R}^{1,2}_5$ and $\widetilde{\mathcal{D}}^{1,2}$.
In the effective theory $\Phi$ is represented by an effective field
\be
\Phi_{\rm eff} = \Phi_0 + a \Phi_1 + \ldots\,,
\ee
where $\Phi_1$ represents a linear combination of O($a$) counterterms 
specific to the field $\Phi$. 
A renormalized lattice correlation function of the field $\Phi$ to order $a$ in the effective theory is 
then given by
\be
\langle \Phi \rangle \rightarrow \langle \Phi_0 \rangle_0 - a \int\!\!d^4y~\langle \Phi_0
     {\mathcal L}_1(y) \rangle_0 + a \langle \Phi_1 \rangle_0 + \ldots
\label{eq:sym_exp}
\ee
where the expectation values $\langle \cdots \rangle_0$ are to be taken in the continuum
theory with action $S_0$.

If we are interested in a non-vanishing correlator in the continuum limit, $\langle \Phi_0 \rangle_0 \neq 0$,
$\Phi_0$ must be even under both $\mathcal{R}^{1,2}_5$ and $\widetilde{\mathcal{D}}^{1,2}$. Because of its 
higher dimensionality, this implies that 
$\Phi_1$ is odd under $\widetilde{\mathcal{D}}^{1,2}$ and thus odd under $\mathcal{R}^{1,2}_5$ as well. We have already noticed
that ${\mathcal L}_1$ is odd under both $\mathcal{R}^{1,2}_5$ and $\widetilde{\mathcal{D}}^{1,2}$.
We can conclude that
both $\langle \Phi_1 \rangle_0$ and $\int\!\!d^4y~ \langle \Phi_0 {\mathcal L}_1(y) \rangle_0$ vanish because the continuum
theory is invariant\footnote{As usual possible contact terms 
in $\int\!\!d^4y~\langle \Phi_0 {\mathcal L}_1(y) \rangle_0$ can be traded for terms with the same symmetry properties 
of $\langle \Phi_1 \rangle_0$ without invalidating the proof.} under $\mathcal{R}^{1,2}_5$.

The key point for the absence of O($a$) terms in the Symanzik expansion of $\mathcal{R}^{1,2}_5$ even correlation
functions is that the continuum action (\ref{eq:tmQCD}) is invariant under 
the discrete chiral transformation $\mathcal{R}^{1,2}_5$,
while all the operators in eq. (\ref{eq:sym_op}) of
the Symanzik expansion of the lattice action are odd under the same discrete chiral symmetry 
transformation. Furthermore, the form of the continuum theory~\eqref{eq:tmQCD}, with vanishing untwisted quark mass,
that guarantees automatic O($a$) improvement is a direct consequence of the non-perturbative tuning of $m_0 = m_{\rm cr}$
in the lattice theory.
Possible uncertainties of O($a$) in the determination of the critical mass $m_{\rm cr}$
are proportional to $\overline{\chi} \chi$, hence their insertions in the effective 
theory vanish as the insertions of ${\mathcal{L}_1}$ do.

Another way of seeing automatic O($a$) improvement is by saying that the lattice action~\eqref{eq:Wtm}
has two distinct sources of chiral symmetry breaking: the Wilson term (together with the critical mass $m_{\rm cr}$)
and the twisted mass term. Automatic O($a$) improvement is a consequence of the fact that 
one of the two terms (the twisted mass term in our basis) retains the discrete chiral symmetry 
$\mathcal{R}^{1,2}_5$. In the next section we show that the same idea applies to the 
$\chi$SF where now the two source of chiral symmetry breaking
are the Wilson term and the boundary conditions satisfied by the fermion fields.

It might come as a surprise that automatic O($a$) improvement works only for correlation
functions that are even under $\mathcal{R}^{1,2}_5$. To understand this we need to do a step 
back to our target continuum theory~\eqref{eq:tmQCD}. We have shown in this section that tmQCD 
and QCD are the same continuum theories written in a different fermion basis. Therefore QCD and tmQCD share 
the same symmetry properties even if the symmetry transformations take different forms in the different
basis. In app.~\ref{app:CPTtwisted} we collect few symmetry transformations in the twisted basis 
for a generic twist angle $\omega$.
The $\mathcal{R}^{1,2}_5$ transformation applied to fermion fields whose
target continuum theory is eq.~\eqref{eq:tmQCD}, i.e. $\omega=\pi/2$, corresponds to a vector (flavor) transformation 
in the basis where the QCD action takes its standard form, i.e. 
the mass term takes its standard form. 
It is easy to see it considering eq.~\eqref{eq:tv} with $\alpha_V^a=(\pi,0,0)$ and 
$\omega=0$ or $\omega=\pi/2$.
To summarize the $\mathcal{R}^{1,2}_5$ transformation takes the form
of a discrete chiral transformation, but it has the physical meaning of a vector (flavor) transformation.
For this reason in the following we will refer to $\mathcal{R}^{1,2}_5$-even correlation functions as ``physical''
to distinguish them from the correlation functions that vanish in the continuum limit.

\section{Chirally rotated Schr\"odinger functional}
\label{sec:chiSF}

A well-known and successful non-perturbative renormalization scheme is the so-called Schr\"odinger
functional (SF) scheme. The SF for QCD is the standard QCD partition function where the fermion
and gauge degrees of freedom satisfy Dirichlet boundary conditions at $x_0=0$ and $T$. Periodic
boundary conditions in the spatial directions are employed for the gauge fields, while fermion fields can be defined to be periodic up to a phase 
\be
\psi(x+L\hat{k}) = {\rm e}^{i \theta_k}\psi(x) \qquad -\pi<\theta_k\le\pi.
\ee
The boundary conditions for the gauge fields are described in detail in 
ref.~\cite{Luscher:1992an,Luscher:1996sc}.
In the following we will concentrate only on the fermionic fields and assume throughout that
the boundary conditions for the gauge fields in the continuum and later on the lattice 
are the standard ones.
A natural choice for Dirichlet boundary conditions (b.c.) for the fermion fields 
are the so-called standard SF b.c.~\cite{Sint:1993un,Luscher:2006df}
\begin{align}
\label{eq:DirichletFermionHomogSpectral}
P_{+} \, \psi(x)\vert_{x_{0} = 0}& =0&
P_{-} \, \psi(x)\vert_{x_{0} = T}& =0 \quad \textrm{from }\mathcal{T}\\
\label{eq:DirichletAntiFermionHomogSpectral}
\overline{\psi}(x) \, P_{-}\vert_{x_{0} = 0}& =0 \quad \textrm{from }\mathcal{C}&
\overline{\psi}(x) \, P_{+}\vert_{x_{0} = T}& =0 \quad \textrm{from
}\mathcal{T}\textrm{ and }\mathcal{C}
\end{align}
with the projectors,
\begin{equation}
  \label{eq:ProjectorParitySFSpectral}
  P_{\pm} = \frac{1}{2}\, \left( 1 \pm \gamma_{0} \right) \,,
\end{equation}
and where we have specified the discrete symmetries (charge conjugation $\mcC$ and time reversal $\mcT$)
that relate the different boundary conditions (see app.~\ref{app:CPTtwisted} for their definitions).
The boundary fermion fields are defined as
\begin{align}
  P_{-} \, \psi(x)\vert_{x_{0} = 0}& =\zeta(\vec{x})&
  P_{+} \, \psi(x)\vert_{x_{0} = T}& =\zeta'(\vec{x}) \label{eq:NonZeroFermionSpectral}\\
  \overline{\psi}(x) \, P_{+}\vert_{x_{0} = 0}& =\overline{\zeta}(\vec{x})&
  \overline{\psi}(x) \, P_{-}\vert_{x_{0} = T}&
  =\overline{\zeta}'(\vec{x}) \, . \label{eq:NonZeroAntiFermionSpectral}
\end{align}

Even in the massless limit standard SF b.c. break chiral symmetry. 
Studying the transformation of the SF propagator under a chiral symmetry transformation one observes that 
the SF b.c. induce a unit mass-like term at the boundaries~\cite{Luscher:2006df}.
The breaking of chiral symmetry by the boundary conditions implies,
contrary to what happens in a finite volume without boundaries,
that Wtm at maximal twist is affected by O($a$) discretization errors. 

From the proof of automatic O($a$) improvement presented in sect.~\ref{sec:Wtm}, we understand that the
relevant property to preserve automatic O($a$) improvement 
is the way chiral symmetry is broken in the continuum theory
(by a mass term or by boundary conditions) with respect to the way the Wilson term does
it in the lattice theory. 
We recall that in the case of Wtm without boundaries the twisted mass term is invariant under 
$\mathcal{R}^{1,2}_5$, while the Wilson term is not.
One might thus think to define the continuum theory with Dirichlet boundary conditions invariant under 
$\mathcal{R}^{1,2}_5$. The presence of the Wilson term at non-zero lattice spacing
should not harm the property of automatic O($a$) improvement~\cite{Sint:2005qz}.

One possible solution to this problem is to mimic exactly what is done with tmQCD.
In the continuum, the boundary conditions preserving $\mathcal{R}^{1,2}_5$
can be obtained from the homogeneous standard SF boundary conditions
via the non-singlet axial transformation defined in eq.~\eqref{eq:twist}.
If we consider a flavor doublet of fermions
and we apply such a rotation to the quark and anti-quark fields, the boundary conditions take the form,
\begin{align}
\label{eq:DirichletFermionHomogQcontSpectrum}
\widetilde{Q}_{+} \, \chi(x)|_{x_{0} = 0}& =0&
\widetilde{Q}_{-} \, \chi(x)|_{x_{0} = T}& =0 \quad \textrm{from
}\mathcal{T}_{\pi/2}\\
\label{eq:DirichletAntiFermionHomogQcontSpectrum}
\overline{\chi}(x) \, \widetilde{Q}_{+}|_{x_{0} = 0}& =0 \quad \textrm{from }\mathcal{C}&
\overline{\chi}(x) \, \widetilde{Q}_{-}|_{x_{0} = T}& =0 \quad
\textrm{from }\mathcal{T}_{\pi/2}\textrm{ and }\mathcal{C}
\end{align}
with projectors
\begin{equation}
\label{eq:QpmFlavorSpectrum}
\widetilde{Q}_{\pm} =
\frac{1}{2} \, \left( 1 \pm \textrm{i}\gamma_{0}\gamma_{5}\tau^{3} \right) \, .
\end{equation}
These are the chirally rotated b.c.~\cite{Sint:2005qz}, which we will refer to as the $\chi$SF b.c.
(see app.~\ref{app:CPTtwisted} for the definition of $\mathcal{T}_{\pi/2}$).
The boundary fields can be defined as
\begin{align}
  \widetilde{Q}_{-} \, \chi(x)\vert_{x_{0} = 0}& =\zeta(\vec{x})&
  \widetilde{Q}_{+} \, \chi(x)\vert_{x_{0} = T}&
  =\zeta'(\vec{x}) \label{eq:NonZeroFermionSpectralQ}\\ 
  \overline{\chi}(x) \, \widetilde{Q}_{-}\vert_{x_{0} = 0}& =\overline{\zeta}(\vec{x})&
  \overline{\chi}(x) \, \widetilde{Q}_{+}\vert_{x_{0} = T}&
  =\overline{\zeta}'(\vec{x}) \, . \label{eq:NonZeroAntiFermionSpectralQ}
\end{align}
It is important to notice that the correspondence between the SF and the $\chi$SF 
is analogous to that between QCD and tmQCD.
The symmetries of the SF are the same as those of QCD
while the symmetries of the $\chi$SF correspond to those
of tmQCD at maximal twist.
In fact, these symmetries are not different in the two formulations, 
they are just expressed in a different basis.
In the continuum, the SF and the $\chi$SF have all the same symmetries.

The transformations of eqs.~\eqref{eq:twist} are a trivial change of basis in the continuum theory,
so one might hope that 
on the lattice massless Wilson fermions with $\chi$SF b.c. will provide a 
framework for a finite volume scheme compatible with automatic O($a$) improvement. 
We have noted before though that standard SF b.c. arise naturally when performing
the continuum limit with massless Wilson fermions. This implies that in order to recover
the $\chi$SF b.c. in the continuum limit 
additional terms have to be added to the Wilson action near the boundaries.
This problem has been solved by Sint in~\cite{Sint:2010eh} using orbifolding techiques.
Orbifolding assures that the proper b.c. are satisfied at tree-level of the lattice theory.
This is sufficient to identify the proper terms to add to the lattice action near the boundaries.
The study of the renormalization of the theory will teach us if additional terms 
are needed to obtain the correct continuum limit in the interacting case.
In the next section we discuss in more details the lattice action proposed in~\cite{Sint:2010eh}. 

We conclude this section by emphasizing that the $\chi$SF 
in the continuum limit is the same renormalization scheme as the standard SF, 
if the same kinematical conditions are chosen. 
It is only at non-zero lattice spacing where the two schemes differ. 
This property is important for many reasons. Here we just mention that 
previous results obtained with the standard SF can be used to check the validity
of the continuum limit of the $\chi$SF. 
We will use this property in our continuum limit scaling studies in
ref.~\cite{Lopez:2012mc}. Moreover, if one is interested in renormalizing
certain operators for which the evolution with the renormalization scale has already been computed
with the standard SF, then it is sufficient to compute the proper
renormalization factors at the lattice spacings where the infinite volume operators 
are used. For the scale evolution the results from the standard SF can then be used.

\section{$\chi$SF with Wilson fermions}
\label{sec:Wilson}

The construction of a lattice action that in the continuum limit goes to QCD with $\chi$SF b.c.
is a non-trivial task.
In ref.~\cite{Sint:2010eh} three lattice actions have been proposed 
which satisfy this property. These actions are the standard Wilson action in the bulk
of the hyper-cylinder with three different local modifications close to the temporal boundaries.
These modifications are necessary to obtain the $\chi$SF b.c. in the continuum limit. 

In this section we briefly discuss the lattice action that we have used in our numerical
investigation. In what follows, we call this the Wilson $\chi$SF (W$\chi$SF) action.
For additional details about the other two formulations see ref.~\cite{Sint:2010eh}.
We consider a doublet of fermions $\chi = \left(\chi_u,\chi_d\right)^{\rm T}$ and a lattice
$L^3 \times \left[0,T\right]$ with spacing $a$.
The W$\chi$SF action reads
\be
\label{eq:FactionWChiSF}
S_{\textrm{F}} =
a^{4} \, \sum_{x_{0}=0}^{T} \sum_{\bx} \, \overline{\chi}(x) \, (
\mathcal{D}_{\textrm{W}} + m_{0} ) \, \chi(x) \,,
\ee
where $x=(x_0,\bx)$ is a point on the lattice with
spatial coordinates $\bx$ and temporal coordinate $x_0$ and
\begin{alignat}{2}
\label{eq:WDopChiSF}
a\mathcal{D}_{\textrm{W}} \, \chi(x)= \,
\begin{cases}
-U_{0}(x)P_{-}\chi(x+a\hat{0}) +
(K + \textrm{i}\gamma_{5}\tau^{3}P_{-})\, \chi(x)
&\textrm{if} \; x_{0}=0\, ,\\
aD_{\textrm{W}} \, \chi(x) &\textrm{if} \; 0 < x_{0} < T\, ,\\
(K + \textrm{i}\gamma_{5}\tau^{3}P_{+})\, \chi(x) -
U_{0}(x-a\hat{0})^{\dagger}P_{+}\chi(x-a\hat{0})
&\textrm{if} \; x_{0}=T\, ,
\end{cases}
\end{alignat}
satisfying, as does the twisted mass operator, the Hermiticity property,
\begin{equation}
\label{eq:g5HermiticityF12DchiSF}
\tau^{1,2} \gamma_{5} \, \mathcal{D}_{\textrm{W}} \, \gamma_{5} \,
\tau^{1,2} = \mathcal{D}_{\textrm{W}}^{\dagger}\, .
\end{equation}
$D_{\textrm{W}}$ is the massless Wilson operator defined in
eq.~\eqref{eq:wilson} and it may be written as
\begin{equation}
\label{eq:WDopInTermsOfK}
aD_{\textrm{W}}\, \chi(x) =
-U_{0}(x)P_{-}\chi(x+a\hat{0}) +
K\chi(x) -
U_{0}(x-a\hat{0})^{\dagger}P_{+}\chi(x-a\hat{0}) \, ,
\end{equation}
with $K$, the dimensionless time-diagonal kernel of the Wilson
operator,
\be
\label{eq:KWilson}
K = 1 + \sum_{k=1}^{3} \, \frac{a}{2} \, \{ \gamma_{k} \, [ \nabla_{k}^{*}(x) +
\nabla_{k}(x) ] - a\, \nabla_{k}^{*}(x) \nabla_{k}(x) \} \, .
\ee
The spectrum of the Hermitean lattice operator,
$\gamma_{5}\tau^{1,2}\mathcal{D}_{\textrm{W}}$,
is bounded from below~\cite{Sint:2010eh}, as in the continuum, 
with a non-vanishing minimum eigenvalue which coincides with the one in the
continuum theory in the limit $a \rightarrow 0$.

We note immediately that the main difference between the W$\chi$SF and the standard SF is the presence of
an additional term 
\be
\chibar(x)\textrm{i}\gamma_{5}\tau^{3}P_{-}\chi(x)\delta_{x_0,0} + \chibar(x)\textrm{i}\gamma_{5}\tau^{3}P_{+}\chi(x) 
\delta_{x_0,T}\,,
\ee
localized at the boundaries.
This term is necessary but not sufficient to recover the proper b.c. in the continuum limit.

To ensure the correct continuum limit, one must account for all
relevant operators allowed by the symmetries of the action above.
This means to consider operators of dimension four or less for the bulk action. There is one
such operator, $\overline{\chi}\chi$, and the corresponding counterterm is the 
term proportional to the critical quark mass, $m_{\rm cr}$.
This is the standard operator that is present for all Wilson actions due to the
breaking of chiral symmetry by the Wilson term.

Similarly, we must include all permitted boundary operators of
dimension three or less. Again, the one allowed operator is
$\overline{\chi}\chi$~\cite{Sint:2010eh}, which gives rise to the following
counterterm to the lattice action,
\begin{displaymath}
\delta S_3 = 
(\zf - 1) a^{3}\sum_{\bx}\, \left( \overline{\chi}\chi|_{x_{0}=0} + \overline{\chi}\chi|_{x_{0}=T} \right)\,.
\end{displaymath}
Such an operator would be forbidden in the continuum action, but the reduced symmetries of the
Wilson action do not allow us to exclude this operator at non-zero lattice spacing.
The presence of $\delta S_3$ can then be understood as necessary to restore in the continuum limit the symmetries 
broken by the Wilson term. 
More specifically the Wilson term, as the $\delta S_3$ term, break the discrete symmetry
$\mathcal{R}_5^{1,2}$. We have noticed in sect.~\ref{sec:Wtm} that to ensure automatic O($a$) improvement 
we want to have a target continuum action and b.c. invariant under an $\mathcal{R}_5^{1,2}$ transformation.
The parameter $\zf$ has to be tuned in order to recover the proper symmetries in the continuum limit
that ensure automatic O($a$) improvement.
Since $\mathcal{R}_5^{1,2}$ is a symmetry of the massless continuum theory,
which is only broken in the regularization procedure,
$\zf$ accounts for a finite renormalization, that is, it has the form,
\begin{equation}
\label{eq:ZfExpansion}
z_{f}(g_{0}) = z_{f}^{(0)} + z_{f}^{(1)} \, g_{0}^{2} + O(g_{0}^{4}) \, ,
\end{equation}
with all coefficients in the expansion being finite.
However, the fact that $\delta S_3$ is not an irrelevant 
operator implies that a perturbative computation of $\zf$ 
is not sufficient and that we then must compute the bare coupling dependence of $\zf(g_0)$ non-perturbatively. 

In perturbation theory, only the tree-level value of $\zf$ is presently known
and it takes the value $z_{f}^{(0)}=1$.
We have determined this value by a direct comparison of the free quark 
propagator in the continuum and the continuum limit of the analytical expression for the free
lattice quark propagator~\cite{jen:2011phd}. The analytical expression of the lattice
tree-level propagator for the action~\eqref{eq:FactionWChiSF} is given in app.~\ref{app:freeWchiSF}.

Furthermore, we must also examine those irrelevant operators that in principle can lead to
O($a$) contributions. In the bulk, there is the dimension five
Sheikholeslami-Wohlert term, but automatic O($a$) improvement eliminates the
need for this operator.
Yet, there does remain an O($a$) contribution from the boundary due to the 
irrelevant dimension four operator~\cite{Sint:2010eh},
\begin{displaymath}
  \delta S_4 = (\ds - 1) a^{4}\sum_{\bx}\,
  \left( \overline{\chi}\gamma_{k}D_{k}\chi|_{x_{0}=0} + \overline{\chi}\gamma_{k}D_{k}\chi|_{x_{0}=T} \right)\,,
\end{displaymath}
where $ D_k = \frac{1}{2}  (\nabla_{k}^{*} + \nabla_{k})$.
Such a contribution is present in all the SF
formulations~\cite{Luscher:2006df} and it is not due to the particular lattice
action or b.c. we have chosen\footnote{In fact, $\ds$ plays a role that is
analogous to the $\tilde{c}_t$ counterterm in the standard
SF~\cite{Luscher:1996sc}.}. Given that $\delta S_4$ is an irrelevant
operator, a perturbative calculation of $\ds$ is presumably sufficient and the expansion
in powers of $g_0^2$ reads
\be
\label{eq:DsExpansion}
d_{s}(g_{0}) = d_{s}^{(0)} + d_{s}^{(1)} \, g_{0}^{2} + O(g_{0}^{4}) \, .
\ee
For the lattice action~\eqref{eq:FactionWChiSF} $d_{s}$ is already needed 
at the tree-level of perturbation theory in order to remove O($a$) boundary cutoff effects.
The tree-level value is for the W$\chi$SF action in eq.~\eqref{eq:FactionWChiSF} $d_{s}^{(0)}=1/2$.
We have determined this value from a numerical inspection of the
free quark propagator on the lattice, obtained from the numerical
inversion of the following lattice Wilson operator 
\be
S_F + \delta S_3 + \delta S_4\,.
\ee
This is in complete agreement with the analytical result obtained in~\cite{Sint:2010eh}.

The knowledge of $d_{s}^{(0)}$ guarantees boundary cutoff effects of at most $O(ag_{0}^{2})$
and like in the standard SF, we expect that a perturbative determination of $d_{s}$ 
is enough to cancel the dominant O($a$) boundary effects. It goes without saying that
a determination beyond tree-level would be very desirable.

The important conclusion of the above discussion is that,
with respect to the standard formulation of the SF,
there is an additional boundary coefficient, $\zf$, which has to be determined
non-perturbatively. However, this is enough to guarantee the correct
continuum limit of the theory and bulk automatic O($a$) improvement
up to boundary effects of at most $O(ag_{0}^{2})$.\footnote{We
note that, similar to the boundary term proportional to $\ds$, 
there is in the gauge action an improvement coefficient $c_t$~\cite{Luscher:1996sc}, 
which multiplies a dimension four boundary term, that is also only known from perturbation theory.}
Therefore, besides the boundary improvement counterterms to the
action, no further improvement counterterms need to be
added to any $\mathcal{R}_5^{1,2}$-even quantity. Thus the
action given in this section retains all the advantages of automatic O($a$) improvement.

\section{Non-perturbative tuning}
\label{sec:tuning}

From purely theoretical considerations, we have concluded in the previous section
that W$\chi$SF provides a suitable discretization for the $\chi$SF
non-perturbative renormalization scheme. 
This is achieved in principle with the non-perturbative tuning of only \emph{two}
parameters which are functions of the bare gauge coupling $g_{0}$: the bare quark mass, $m_{0}$, 
and the boundary coefficient, $\zf$. The bare quark mass needs to be tuned to its critical value, $\mcr$,
in order to have a massless scheme, while the tuning of the coefficient $\zf$ to its critical
value, $\zfcr$, is required in order to recover the desired boundary conditions in the
continuum and thus to obtain bulk automatic O($a$) improvement.

If $\zfcr$ is not determined correctly, then the $\mathcal{R}_5^{1,2}$-symmetry would not be properly 
restored and bulk automatic O($a$) improvement would not take place.  
The loss of automatic O($a$) improvement may be the least of our worries 
if the continuum limit itself is compromised by incorrectly fixing $\zfcr$. 
Clearly, a non-perturbative determination of $\zfcr$ is then mandatory.

The non-perturbative determination of $\zfcr$ can be carried out by requiring suitable
$\mathcal{R}_5^{1,2}$-odd correlation functions to vanish (cf. also~\cite{Sint:2010xy}).
Since these conditions are not unique, different determinations of $\zfcr$ are expected to differ by O($a$) effects,
which should only affect $\mathcal{R}_5^{1,2}$-even correlation functions up to $O(a^{2})$.
This is similar to what happens in large volume simulations with Wilson
twisted mass fermions at maximal twist. The intrinsic O($a$) uncertainties in
the determination of the critical mass $\mcr$ only affect physical quantities at
$O(a^{2})$.

In fact both $m_{0}$ and $\zf$
have to be tuned non-perturbatively and simultaneously if a massless
renormalization scheme with $\chi$SF boundary conditions is to be
defined. 
In particular, it is very important to understand whether this
`combined' tuning is feasible at all, as otherwise a practical
application of the $\chi$SF scheme would be rather cumbersome.

After the proper determination of $\mcr$, $\zfcr$
and the two boundary improvement coefficients to the action, $c_{t}$
and $d_{s}$, automatic O($a$)-improvement is expected to hold.
This means that without any improvement counterterm to the action in
the bulk and to the fields, all physical quantities have leading $O(a^{2})$ discretization effects.
In practice, only a perturbative determination of the boundary
improvement coefficients, $c_{t}$ and $d_{s}$, is available.
At present, only the tree-level value of $d_{s}$ is known
in perturbation theory. For $c_{t}$, we employ the 2-loop value~\cite{Bode:1998hd,Bode:1999sm},
$c_{t}(g_{0}) = 1 - 0.089 \, g_{0}^{2} - 0.030 \, g_{0}^{4}$.

Here we are concerned with the non-perturbative tuning of
the other two coefficients, $m_{0}$ and $\zf$.
From now on, as it is usually done, all discussions will take
place in terms of the hopping parameter $\kappa = \frac{1}{8+2am_{0}}$.
Due to the potential complications which may arise in the tuning procedure,
we first performed some studies at the tree-level of perturbation theory.
We have tested several tuning strategies, and 
the preferred one, as it emerged from our tree-level investigation,
was applied non-perturbatively to the interacting theory~\cite{Lopez:2009yc} 
as will be explained in sect.~\ref{sec:Tuning.TheStrategy}.
Besides the particular selection of the tuning strategy,
a tuning condition must also be chosen. In sect.~\ref{ssec:Tuningconditions} we will
describe all the tuning conditions we have investigated.

\subsection{Some definitions}
\label{ssec:somedef}

The non-perturbative determination of $\kcr$ and $\zfcr$
requires imposing conditions at non-zero lattice spacing that ensure the
restoration of the expected symmetries in the continuum limit
that are broken by the Wilson term at non-zero lattice spacing.
Moreover, these conditions should be imposed at each lattice spacing
while fixing a suitable renormalized quantity. In this work, we keep
the renormalized SF coupling, $\overline{g}$, fixed.  This is
equivalent to fixing the physical size of the box, $L$ (we choose $T=L$).
All other external parameters must also be held fixed. These are $d_{s}$, which is set to its 
tree-level value, $d_{s}^{(0)}$, and the spatial momentum, $\bp$, which is set to
zero.
In the spatial directions, periodic boundary conditions up to a
phase are assumed, whose phase dependence is parametrized by the
angles $\boldsymbol{\theta}=(\theta_{1},\theta_{2},\theta_{3})$.
During the tuning procedure, these angles are used
in order to define alternative tuning conditions.
To be concrete, we choose the symmetric case, $\theta_{k}=\theta$
($k=1,2,3$) and two values of $\theta$ are used, $\theta = \theta_{A}=0$ and
$\theta = \theta_{B}=0.5$. Different choices for $\theta$ are used so that 
we can define tuning conditions that differ by O($a$)
The choices for the external parameters are summarized in
tab.~\ref{tab:FixedParameters}. 
\begin{table}
\centering
\begin{tabular}[c]{| c | c | c | c | c | c | c | c | c |}
\hline
$T/L $  & $d_{s}$ & $p_{1}$ &  $p_{2}$ & $p_{3}$ & $\theta_{A}$ &
$\theta_{B}$ & $\overline{x}_{0}$ & $\overline{y}_{0}$ \\
\hline
1     &  0.5 &  0.0  & 0.0 & 0.0 & 0.0  & 0.5 & T/2 & 3T/4 \\
\hline
\end{tabular}
\caption{Fixed parameters during the tuning. See the text for a discussion of these parameters.}
\label{tab:FixedParameters}
\end{table}

Before specifying the tuning conditions,
we define the correlation functions that are needed for our tuning
procedure. In particular, we will employ boundary to bulk correlation
functions that involve the boundary at $x_{0}=0$.
For this purpose, we first define the boundary
operators.
A definition of the fermion fields at the boundary ($x_0=0$) consistent with gauge invariance is given by
\be
\label{eq:DefZetaFieldsTuning}
\zeta({\bx}) = U_{0}(0,{\bx})\chi(a,\bx)
\qquad
\overline{\zeta}({\bx}) = \overline{\chi}(a,\bx)U_{0}(0,{\bx})^{\dagger} \, .
\ee
The boundary interpolating fields at $x_{0}=0$ are given by
\begin{equation}
\label{eq:bopXTun}
\widetilde{\mathcal{O}}_{\pm}^{a} = a^{6}\sum_{\by,\bz} 
\overline{\zeta}(\by)\Gamma_{\widetilde{\mathrm{O}}}\widetilde{Q}_{\pm}\zeta(\bz)\, .
\end{equation}
In this expression, $\widetilde{Q}_{\pm}$ are the $\chi$SF projectors
defined in sect.~\ref{sec:chiSF}.
$\Gamma_{\widetilde{\mathrm{O}}}$ contains the flavor and Dirac structure
of an operator of type $\widetilde{O}$. Specifically
for a pseudo-scalar density and an axial-vector current, we have
\begin{subequations}
\label{eq:bopTun}
\begin{align}
\widetilde{\mathcal{P}}_{\pm}^{a} &= a^{6}\sum_{\by,\bz} 
\overline{\zeta}(\by)\gamma_{5}\frac{\tau^{a}}{2}\widetilde{Q}_{\pm}\zeta(\bz)\,
, \label{eq:bopTun_P}\\
\widetilde{\mathcal{A}}_{\mu \pm}^{a} &= a^{6}\sum_{\by,\bz} 
\overline{\zeta}(\by)\gamma_{\mu}\gamma_{5}\frac{\tau^{a}}{2}\widetilde{Q}_{\pm}\zeta(\bz)\,
. \label{eq:bopTun_Amu}
\end{align}
\end{subequations}
We note that there is a little difference with respect to the SF
formulation, where the projectors are included in the
definition of the boundary fields, $\zeta, \overline{\zeta}$.
Here we insert the projectors directly in the correlations functions
to have the freedom to consider correlation functions with the ``wrong'' projectors~\cite{Sint:2010xy}.
These correlation functions ought to vanish in the continuum limit if the correct
$\chi$SF b.c. are recovered, and this is confirmed numerically (see sect.~\ref{ssec:R512odd}).

Considering the previous definitions of the boundary interpolating
fields, we may introduce now our notation for the boundary to bulk
correlation functions.
Given a bulk operator, $X^{a}(x)$, the type of correlation functions
that we consider here are the following,
\begin{subequations}
\label{eq:bvcfGeneral}
\begin{align}
g_{\mathrm{X}_{\pm}}^{ab}(x_{0}, \theta) &=
-\frac{a^{3}}{L^{3}}\sum_{\bx}\langle X^{a}(x)\widetilde{\mathcal{P}}_{\pm}^{b}\rangle \label{eq:bvcf_g} \, , \\
\overline{g}_{\mathrm{X}_{\pm}}^{ab}(x_{0}, \theta) &=
-\frac{a^{3}}{L^{3}}\sum_{\bx}\langle X^{a}(x)\widetilde{\mathcal{A}}_{\mu
  \pm}^{b}\rangle \label{eq:bvcf_gBar} \, .
\end{align}
\end{subequations}
For the tuning we used only the particular cases,
\begin{subequations}
\label{eq:bvcfTun}
\begin{align}
g_{\mathrm{P}_{\pm}}^{ab}(x_{0}, \theta) &= -\frac{a^{3}}{L^{3}}\sum_{\bx}\langle
P^{a}(x)\widetilde{\mathcal{P}}_{\pm}^{b}\rangle \label{eq:bvcfTun_P} \, , \\
g_{\mathrm{A}_{\mu \pm}}^{ab}(x_{0}, \theta) &= -\frac{a^{3}}{L^{3}}\sum_{\bx}\langle
A_{\mu}^{a}(x)\widetilde{\mathcal{P}}_{\pm}^{b}\rangle \label{eq:bvcfTun_A}
\, ,\\
\overline{g}_{\mathrm{V}_{\mu \pm}}^{ab}(x_{0}, \theta) &= -\frac{a^{3}}{L^{3}}\sum_{\bx}\langle
V_{\mu}^{a}(x)\widetilde{\mathcal{A}}_{\mu
  \pm}^{b}\rangle \label{eq:bvcfTun_V} \, .
\end{align}
\end{subequations}
We denote the correlation functions in this work with $g_{\mathrm{X}}$, in order to 
distinguish them from the corresponding correlation functions in the standard SF
usually denoted with $f_{\mathrm{X}}$. One may interprete the $g_{\mathrm{X}}$ as correlators in the
$\chi$-basis, while the $f_{\mathrm{X}}$ refer to the standard basis.
The superscripts, $a,b$, denote the flavor index.
The subscripts $X$, in $g_{\mathrm{X}}$, indicate the 
corresponding operator inserted in the bulk of the lattice.
Three bulk operators are considered here:
the pseudo-scalar density, $P^{a}(x)$,
and the axial-vector and vector currents,
$A_{\mu}^{a}(x)$, $V_{\mu}^{a}(x)$.
Depending on what $\chi$SF projector, $\widetilde{Q}_{\pm}$, is chosen in the correlation functions,
we have the corresponding subscript $\pm$.
Due to the particular $\chi$SF boundary conditions, cf.
Eq.~\eqref{eq:DirichletFermionHomogQcontSpectrum}-\eqref{eq:DirichletAntiFermionHomogQcontSpectrum},
all correlation functions defined through $\widetilde{Q}_{+}$ at $x_{0}=0$ should
vanish in the continuum limit. The same holds for $\widetilde{Q}_{-}$ at
$x_{0}=T$. However, we do not consider here such correlation functions.
Therefore, such kind of correlation functions will
be used only later on to perform checks on the recovery of the correct b.c. in the continuum limit. 
For the tuning conditions we consider only correlation functions defined through $\widetilde{Q}_{-}$.

As a last consideration before going into the details of the particular
tuning conditions, we also need to define the correlation function,
\begin{equation}
\label{eq:GA_def}
G_{\mathrm{A}_{\mu \pm}}^{ab}(x_{0}, y_{0} ; \theta, \theta') \equiv
(g_{\mathrm{I}})_{\mathrm{A}_{\mu \pm}}^{ab}(x_{0}, \theta) - s(x_{0}, \theta)\,
\frac{(g_{\mathrm{I}})_{\mathrm{A}_{\mu \pm}}^{ab}(y_{0}, \theta) -
  (g_{\mathrm{I}})_{\mathrm{A}_{\mu \pm}}^{ab}(y_{0}, \theta')
}{s(y_{0}, \theta) - s(y_{0},\theta')} \, .
\end{equation}
The notation is the following.
Let us consider the improved axial current,
\begin{equation}
\label{eq:ACImprChiSFTuning}
(A_{\mathrm{I}})_{\mu}^{a}(x) = A_{\mu}^{a}(x)
+
a\, c_{\mathrm{A}} \, \widetilde{\partial}_{\mu} \, P^{a}(x) \, ,
\end{equation}
where the derivative on the lattice, $\widetilde{\partial}_{\mu}$, is
defined to be the symmetric derivative,
\begin{equation}
\label{eq:DefSymPartDer}
\widetilde{\partial}_{\mu} \equiv \frac{1}{2}\,
(\partial_{\mu}^{*}+\partial_{\mu}) \, ,
\end{equation}
with the standard definition of the partial derivatives on the lattice.
The correlation function
$(g_{\mathrm{I}})_{\mathrm{A}_{\mu \pm}}^{ab}(x_{0}, \theta)$
is defined as
\begin{equation}
  \label{eq:bvcfTun_AImp}
  (g_{\mathrm{I}})_{\mathrm{A}_{\mu \pm}}^{ab}(x_{0}, \theta) =
  -\frac{a^{3}}{L^{3}}\sum_{\bx}\langle
  (A_{\mathrm{I}})_{\mu}^{a}(x)\widetilde{\mathcal{P}}_{\pm}^{b}\rangle
  \, .
\end{equation}
This is just the equivalent of Eq.~\eqref{eq:bvcfTun_A}, where the
expression of the improved axial current is used instead of the
unimproved one. Eq.~\eqref{eq:bvcfTun_AImp} may be rewritten in terms of
Eq.~\eqref{eq:bvcfTun_P}  and Eq.~\eqref{eq:bvcfTun_A} as follows,
\begin{equation}
\label{eq:bvcfTun_AImp2}
(g_{\mathrm{I}})_{\mathrm{A}_{\mu \pm}}^{ab}(x_{0}, \theta) =
g_{\mathrm{A}_{\mu \pm}}^{ab}(x_{0},\theta)
+ c_{\mathrm{A}} \, s(x_{0}, \theta) \, ,
\qquad
s(x_{0},\theta) \equiv a \, \widetilde{\partial}_{\mu}\,
g_{\mathrm{P}_{\pm}}^{ab}(x_{0}, \theta) \, .
\end{equation}
By substitution of Eq.~\eqref{eq:bvcfTun_AImp2} into Eq.~\eqref{eq:GA_def},
the last can be cast in a more explicit manner,
\begin{equation}
\label{eq:GA_implemented}
\begin{split}
G_{\mathrm{A}_{\mu \pm}}^{ab}(x_{0},y_{0};\theta,\theta') &=
g_{\mathrm{A}_{\mu \pm}}^{ab}(x_{0}, \theta)\\
&- [g_{\mathrm{A}_{\mu \pm}}^{ab}(y_{0},
\theta) - g_{\mathrm{A}_{\mu \pm}}^{ab}(y_{0}, \theta')]\,
\frac{\widetilde{\partial}_{\mu}g_{\mathrm{P}_{\pm}}^{ab}(x_{0},
  \theta)}{\widetilde{\partial}_{\mu}g_{\mathrm{P}_{\pm}}^{ab}(y_{0},
  \theta) - \widetilde{\partial}_{\mu}g_{\mathrm{P}_{\pm}}^{ab}(y_{0},
  \theta')}\,.
\end{split}
\end{equation}
This expression is independent of the improvement coefficient
of the axial current, $c_{\mathrm{A}}$, and it
indicates that, $G_{\mathrm{A}_{\mu \pm}}^{ab}(x_{0},y_{0};\theta,\theta')$
equals $g_{\mathrm{A}_{\mu \pm}}^{ab}(x_{0}, \theta)$ up to cutoff effects of
leading O($a$). 

The correlation function $G_{\mathrm{A}_{\mu \pm}}^{ab}(x_{0},y_{0};\theta,\theta')$ is useful
because it allows us to define yet another tuning condition differing by O($a$) effects. 
In particular, in the way it is constructed, this correlation function would not depend on the cutoff effects
that in the standard setup are removed by employing a non-perturbatively tuned $c_A$ improvement coefficient.

\subsection{Tuning conditions}
\label{ssec:Tuningconditions}

As already anticipated, imposing distinct symmetry restoration conditions would
give rise to different values of $\kcr$ and $\zfcr$
due to cutoff effects.
Therefore, it is important to study the sensitivity of
$\kappa$ and $\zf$ to the particular definitions
in order to better understand the intrinsic uncertainty
in the determination of these counterterms. Our exploratory studies~\cite{Lopez:2009yc} have shown that
the tuning of $\kappa$ does not pose any special problem in the $\chi$SF
setup with respect to the standard formulation with Wilson fermions, 
thus we have decided to investigate only one tuning condition for $\kappa$.
On the contrary, as it will be shown later, there is a rather big
sensitivity on the choice of the tuning condition used to
determine $\zfcr$.
Therefore, we have concentrated our efforts in the investigation
of different tuning conditions for $\zf$,
where we have studied seven different possibilities,
which we number from (1) to (7).
Note that even if different tuning conditions have been used to
define $\zfcr$, the tuning strategy we have employed is the same
for all the tuning conditions (cf. sect.~\ref{sec:Tuning.TheStrategy}).

To tune $\kappa$ to its critical value we adopt the standard procedure
of imposing a vanishing PCAC mass. To be concrete, it is defined here as,
\begin{equation}
\label{eq:TuningConditionKappa}
m_{\mathrm{PCAC}} \equiv \frac{\widetilde{\partial}_{0}
  g_{\mathrm{A}_{0 -}}^{11}(\overline{x}_{0},\theta_{A})}{2g_{\mathrm{P}_{-}}^{11}(\overline{x}_{0},\theta_{A})} \, .
\end{equation}
To tune $\zf$ we require a $\mathcal{R}_5^{1,2}$-odd correlation function
to vanish.
The correlation functions (from (1) to (7)) that we have used to tune
$\zf$ are the following,
\begin{subequations}
\label{eq:tuning_conditions_zf}
\begin{align}
(1) &\equiv g_{\mathrm{A}_{0 -}} \equiv g_{\mathrm{A}_{0 -}}^{11}(\overline{x}_{0}, \theta_{A})\,,  \label{eq:tuning_conditions_zf_1}\\
(2) &\equiv g'_{\mathrm{A}_{0 -}} \equiv g_{\mathrm{A}_{0 -}}^{11}(\overline{x}_{0}, \theta_{B})\,,
\label{eq:tuning_conditions_zf_2}\\
(3) &\equiv g_{\mathrm{A}_{0 -}}^{\mathrm{diff}} \equiv
g_{\mathrm{A}_{0 -}} - g'_{\mathrm{A}_{0 -}} \,,\label{eq:tuning_conditions_zf_3}\\
(4) &\equiv \overline{g}_{\mathrm{V}_{k -}} \equiv \frac{1}{3}\,\sum_{k=1}^{3} \, \overline{g}_{\mathrm{V}_{k
    -}}^{12}(\overline{x}_{0}, \theta_{A}) \,,\label{eq:tuning_conditions_zf_4}\\
(5) &\equiv \overline{g}'_{\mathrm{V}_{k -}} \equiv \frac{1}{3}\,\sum_{k=1}^{3} \, \overline{g}_{\mathrm{V}_{k
    -}}^{12}(\overline{x}_{0}, \theta_{B}) \,, \label{eq:tuning_conditions_zf_5}\\
(6) &\equiv \overline{g}_{\mathrm{V}_{k -}}^{\mathrm{diff}} \equiv
\overline{g}_{\mathrm{V}_{k -}} - \overline{g}'_{\mathrm{V}_{k -}}\,, \label{eq:tuning_conditions_zf_6}\\
(7) &\equiv G_{\mathrm{A}} \; \, \, \equiv G_{\mathrm{A}_{0
    -}}^{11}(\overline{x}_{0}, \overline{y}_{0} ; \theta_{A}, \theta_{B}) \,. \label{eq:tuning_conditions_zf_7}
\end{align}
\end{subequations}
The values of the parameters used in the
eq.~\eqref{eq:TuningConditionKappa} and eq.~\eqref{eq:tuning_conditions_zf}
are given in tab.~\ref{tab:FixedParameters}.
In all these conditions,
the particular combinations of interpolating fields
with their corresponding Dirac and flavor indices
are chosen such that the resulting
correlation function is $\mathcal{R}_5^{1,2}$-odd and non-vanishing by definition at non-zero lattice spacing.
That is, the correlation functions do not violate any
symmetry of the lattice theory.
Eq.~\eqref{eq:TuningConditionKappa} and the conditions (1)-(6)
are obtained directly from the definitions in Eq.~\eqref{eq:bvcfTun}
with the corresponding substitutions.
The condition (7) is obtained from Eq.~\eqref{eq:GA_def}, also with
the corresponding substitutions.

All the different tuning conditions introduced in this section
allow us to study the dependence of the 'physical' correlation functions on the 
different values of $\zfcr$ obtained. 
In addition, having a number of tuning conditions at our disposal 
enables us to test the universality of the continuum limit.
 
In this work,
as is the usual choice in SF schemes,
we have defined the correlation functions in the middle of the
time-extent of the lattice, $\bar{x}_{0}=T/2$. The only exception is
the condition (7), which involves two time slices. There the choice is
$\bar{x}_{0}=T/2$ and $\bar{y}_{0}=3T/4$. The reason for these choices is to
stay as far away as possible from the boundaries, thus avoiding boundary
effects. In case of (7), the condition was to maximize the distance between the 
two time-slices, while still staying away from the boundaries. 

Before turning to our tuning strategy, we need to make one
last remark regarding our particular choices of tuning conditions.
In order to restore the symmetries of the theory, we impose that the
different symmetry-violating correlation functions vanish at non-zero
lattice spacing. However, to remove cutoff effects at tree-level, 
a better choice would be to force the corresponding correlation function to take its tree-level value
at non-zero lattice spacing.
From our studies at tree-level~\cite{jen:2011phd},
we have seen that such effects are very small, well below our statistical
accuracy, and do not change our final results.

\subsection{Tuning strategy}
\label{sec:Tuning.TheStrategy}

To check the viability of the tuning strategy for $\kappa$ and $\zf$
non-perturbatively, we have performed a tuning at three
values of the renormalization scale $\mu=1/L$ using only method (1).
This situation corresponds to the results presented
in~\cite{Lopez:2009yc}, where we first explained our tuning strategy
at the non-perturbative level. This separate analysis has been useful
to check the tuning procedure. Results obtained by this analysis are
labelled here as obtained with method (1*).
The three scales correspond to a
hadronic ($\overline{g}^{2}$ fixed with $L = 1.436\, r_{0}$), an
intermediate ($\overline{g}^{2}=2.4484$) and a perturbative
($\overline{g}^{2}=0.9944$) scale.

In sect.~\ref{ssec:Tuningresults} we will present results obtained following the strategy 
presented in this section for all the other tuning conditions
defined in sect.~\ref{ssec:Tuningconditions}.

\begin{figure}
  \centering
  \includegraphics[width=0.7\textwidth]{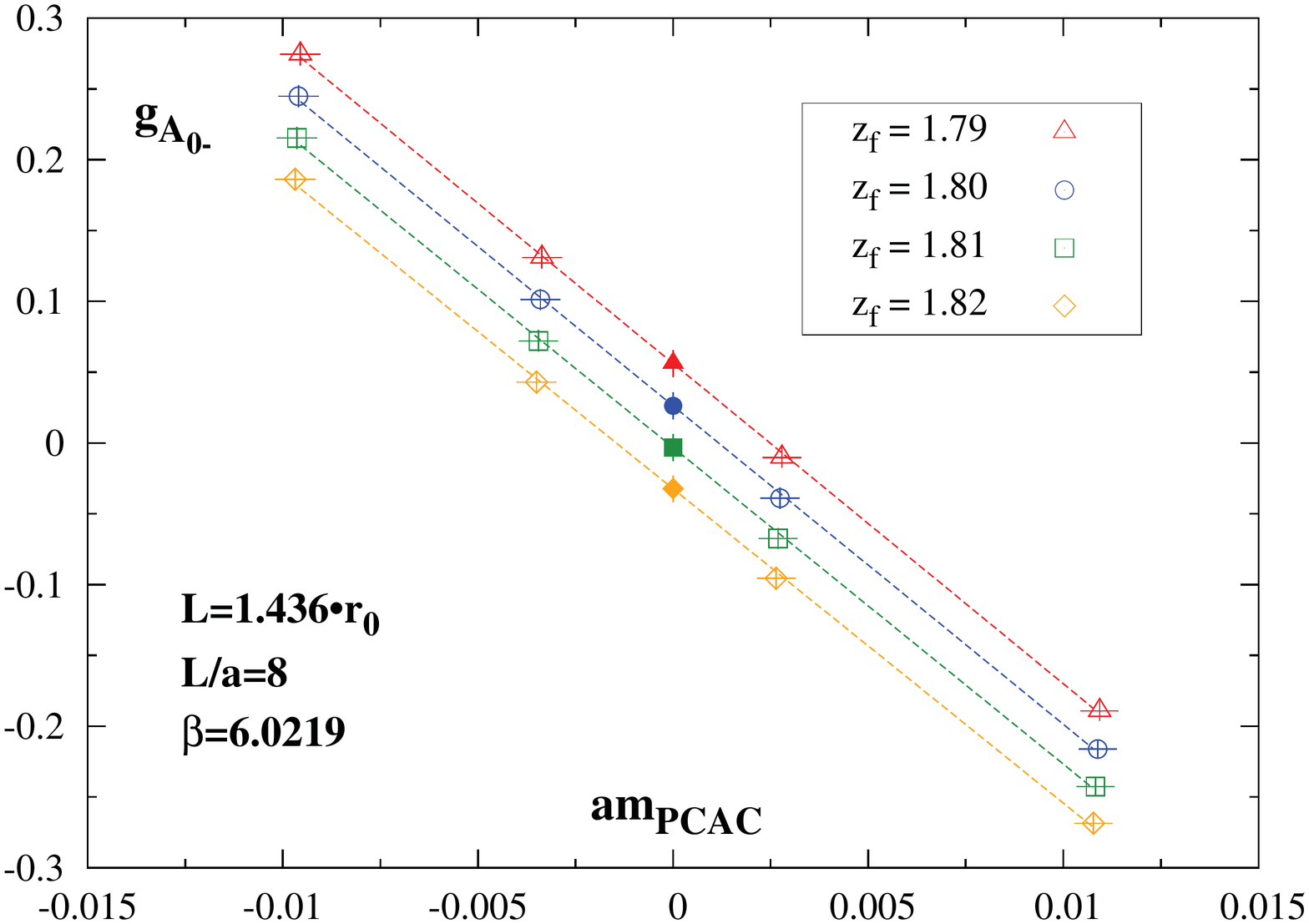}
  \caption{$g_{\mathrm{A}_{0 -}}$ vs. $am_{\mathrm{PCAC}}$
    at four values of $\zf$ (open symbols).
    All fits are linear in $am_{\mathrm{PCAC}}$.
    The values of $g_{\mathrm{A}_{0 -}}$ at
    $am_{\mathrm{PCAC}}=0$,
    denoted $g_{\mathrm{A}_{0 -}}^{*}$, are also
    plotted (filled symbols).}
  \label{fig:ga_mpcac}
\end{figure}

\begin{figure}
  \centering
  \includegraphics[width=0.7\textwidth]{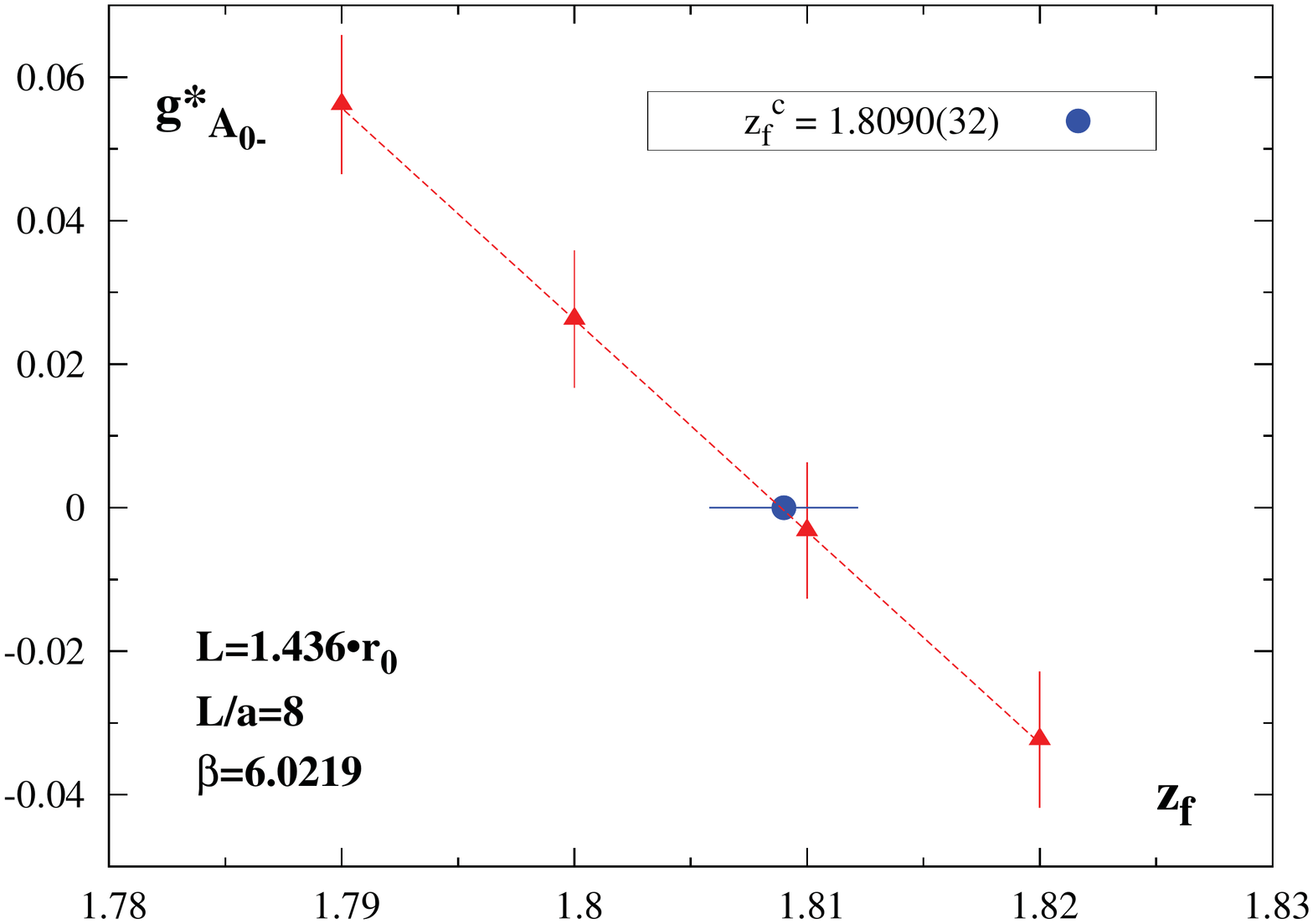}
  \caption{$g_{\mathrm{A}_{0 -}}^{*}$ vs. $\zf$ (red).
    The fit is linear in $\zf$.
    The value of $\zfcr$, $\zf(g_{\mathrm{A}_{0 -}}^{*}=0)$,
    is also plotted (blue).}
  \label{fig:ga_zf}
\end{figure}

The same tuning strategy has been used for each value of $\beta$ and the corresponding value 
of $L/a$.
The values of $\beta$ used are given in tabs.~\ref{tab:tuning_guess_np_new}-\ref{tab:tuning_guess_p_new} 
and are taken from~\cite{Capitani:1998mq}.
The tuning is performed in several steps.
\begin{itemize}
\item We calculate $am_{\mathrm{PCAC}}$ and $g_{\mathrm{A}_{0 -}}$ at four values of
$\zf$, and for each value of $\zf$, we use four values of $\kappa$, thus
giving 16 pairs of $\kappa$ and $\zf$.  This allows us to determine
$g_{\mathrm{A}_{0 -}}$ as a function of $am_{\mathrm{PCAC}}$ for each value of $\zf$, as
illustrated in fig.~\ref{fig:ga_mpcac}.
\item For each value of $\zf$, we perform a linear interpolation of
$g_{\mathrm{A}_{0 -}}$ in terms of $am_{\mathrm{PCAC}}$ to the point $am_{\mathrm{PCAC}}=0$.
This determines the values of $g_{\mathrm{A}_{0 -}}$ at $am_{\mathrm{PCAC}}=0$,
denoted $g_{\mathrm{A}_{0 -}}^{*}$,
for each of the four values of $\zf$,
as shown in fig.~\ref{fig:ga_mpcac} as the filled symbols.
\item We now interpolate these values
of $g_{\mathrm{A}_{0 -}}^{*}$ as a function of $\zf$ to the point of vanishing
$g_{\mathrm{A}_{0 -}}^{*}$,
thus giving us the critical value $\zfcr$, as shown in fig.~\ref{fig:ga_zf}.
\end{itemize}

\begin{figure}[t]
\centering
\includegraphics[width=0.7\textwidth]{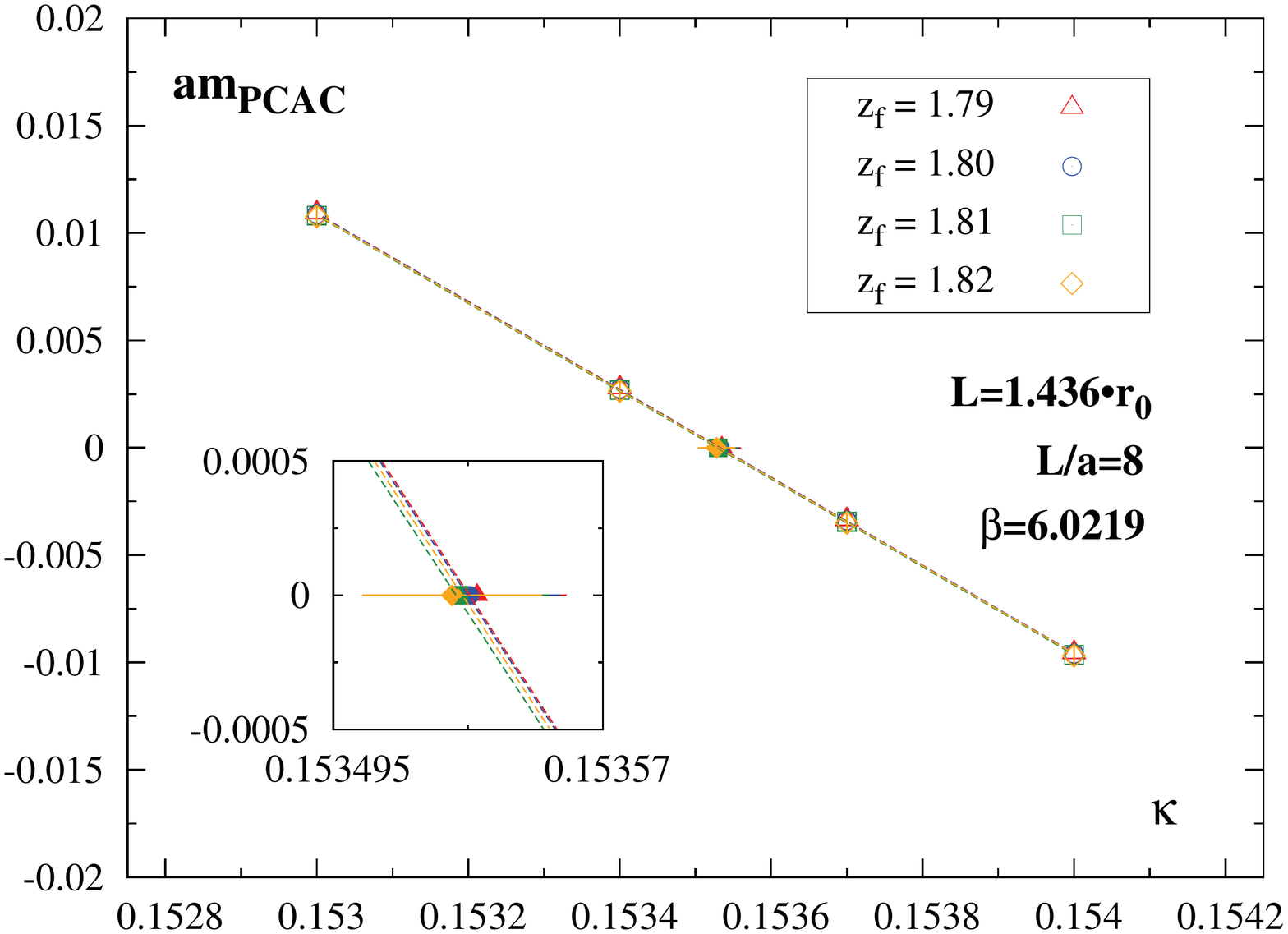}
\caption{$am_{\mathrm{PCAC}}$ vs. $\kappa$
  at four values of $\zf$ (open symbols).
  All fits are linear in $\kappa$.
  The values of $\kappa$ at
  $am_{\mathrm{PCAC}}=0$, denoted $\kappa^{*}$, are also plotted
  (filled symbols).}
\label{fig:mpcac_kappa}
\end{figure}

\begin{figure}
\centering
\includegraphics[width=0.7\textwidth]{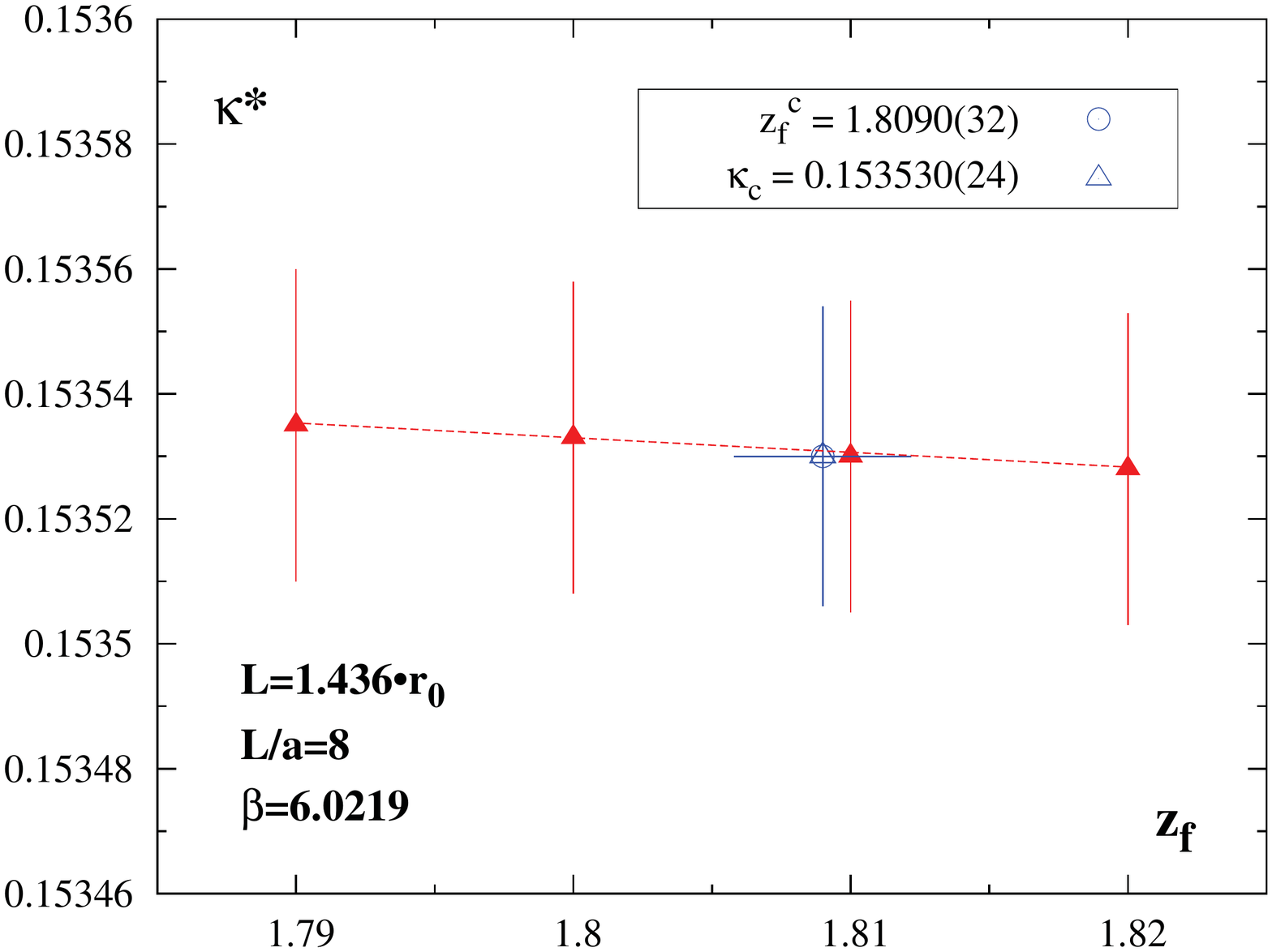}
\caption{$\kappa^{*}$ vs. $\zf$ (red).
  The fit is linear in $\zf$.
  The value of $\zfcr$ (cf. fig.~\ref{fig:ga_zf})
  and of $\kcr$, defined as $\kappa^{*}(\zf=\zfcr)$,
  are also plotted (blue).}
\label{fig:kappa_zf}
\end{figure}

All the numerical data for these intermediate steps can be found in ref.~\cite{jen:2011phd}.
Next we determine $\kcr$.  
\begin{itemize}
\item Using the same 16 pairs of $\kappa$ and
$\zf$, we calculate $am_{\mathrm{PCAC}}$ as a function of $\kappa$ for each $\zf$.
This is shown in fig.~\ref{fig:mpcac_kappa}.
Note that $am_{\mathrm{PCAC}}$ has a very mild dependence on $\zf$, so the four curves at fixed $\zf$
are nearly indistinguishable.  Interpolating $am_{\mathrm{PCAC}}$ in $\kappa$ to the point of
vanishing PCAC mass, $\kappa^{*}$, we obtain the values of $\kappa^{*}$ at each $\zf$.
The resulting values of $\kappa^{*}$ as a function of $\zf$ are
shown in fig.~\ref{fig:kappa_zf}.
\item We now interpolate these results in $\zf$ to the previously determined
value of $\zfcr$, thus determining the value of $\kcr$.
\end{itemize}

All the tuning results can be read off from tab.~\ref{tab:kc_beta} and tab.~\ref{tab:zfc_beta}.

A key observation of this work is the mild dependence of $am_{\mathrm{PCAC}}$ on
$\zf$, at least in the region near $\kcr$ and $\zfcr$.
This can be easily seen in fig.~\ref{fig:mpcac_kappa}.
The consequence of this is clear in fig.~\ref{fig:kappa_zf}:
the determination of $\kcr$ has a weak dependence on $\zfcr$
and the errors of both of them are relatively independent.

\subsection{Tuning results}
\label{ssec:Tuningresults}

With the strategy discussed in sect.~\ref{sec:Tuning.TheStrategy},
we have performed the tuning of $\kappa$ and $\zf$ using
$m_{\mathrm{PCAC}}$ in Eq.~\eqref{eq:TuningConditionKappa}
and the conditions (1)-(7) defined in
Eq.~\eqref{eq:tuning_conditions_zf}.
The tuning has been performed
for five fixed values of the renormalized gauge coupling,
$\overline{g}(L)$, which correspond to five values of the physical
energy scale, $1/L$.
In particular, the physical scale ranges from the purely
non-perturbative to the perturbative regime and 14 values of $\beta$
have been considered within that range.
For better clarity, the tuning points are summarized in
tab.~\ref{tab:TuningPoints}.
The notation in this table is the following. `Scale' refers to the
physical scale, namely, the fixed value of the renormalized gauge
coupling. We have denoted the five scales as `NP', `I', `P', `2P' and
`PP', from the hadronic to the most perturbative scale.
NP corresponds to $L = 1.436\, r_{0}$, I to $\overline{g}^{2}=2.4484$ and
P to $\overline{g}^{2}=0.9944$.
For 2P and PP we have not determined the gauge coupling explicitly.
These two scales have been considered in order to study the dependence
of $\zf$ on $g_{0}$, for small values of $g_{0}$,
thinking of a future perturbative determination of $\zf$,
for which the knowledge of the renormalized coupling is not
necessary.
The results obtained for $\kcr$ and $\zfcr$ using the strategy outlined in sect.~\ref{sec:Tuning.TheStrategy}
for all the tuning methods are summarized 
in tab.~\ref{tab:kc_beta} and tab.~\ref{tab:zfc_beta} for
$\kcr$ and $\zfcr$, respectively.
We also present tables showing the values used for $\zf$
and $\kappa$, at each of the points where the tuning was performed.
In these same tables, the column labelled `$N_{\textrm{conf}}$'
represents the number of configurations used in the computation of all
observables at the corresponding point. These are
tab.~\ref{tab:tuning_guess_np_new} to tab.~\ref{tab:tuning_guess_pp_new}.

\begin{table}
\centering
\begin{tabular}[c]{| c | r | r | r |}
\hline
\multicolumn{1}{|c|}{ Scale} &
\multicolumn{1}{|c|}{$L/a$} &
\multicolumn{1}{|c|}{$\beta$} &
\multicolumn{1}{|c|}{ Name} \\
\hline
$L = 1.436\,r_{0}$ &  8 & 6.0219 & NP \\
                         & 10 & 6.1628 &\\
                         & 12 & 6.2885 &\\
                         & 16 & 6.4956 &\\
                         & 20 & 6.6790 &\\
                         & 24 & 6.8187 &\\
\hline
$\overline{g}^{2}=2.4484$ &   8 & 7.0197  & I \\
                                          & 12 & 7.3551 &  \\
                                          & 16 & 7.6101 &  \\
\hline
$\overline{g}^{2}=0.9944$ &   8 & 10.3000 & P \\
                                          & 12 & 10.6086 & \\
                                          & 16 & 10.8910 & \\
\hline
            & 16 & 12.0000 & 2P \\
\hline
             & 16 & 24.0000 & PP \\
\hline
\end{tabular}
\caption{Summary of all the points where the tuning was performed.}
\label{tab:TuningPoints}
\end{table}

During the tuning,
we have used several combinations of $\kappa$
and $\zf$.
As indicated in sect.~\ref{sec:Tuning.TheStrategy}, the usual
choice is to use 4 values of $\kappa$ and 4 values of $\zf$.
However, there are cases where we have used, instead, 5 values of
$\zf$ and/or 2 values of $\kappa$.
In particular, we have used 2 values of $\kappa$ at all the $\beta$
values where we also performed the separate tuning using (1*).
The reason is that, relying on the very weak dependence of $\kappa$
and $\zf$ on each other, as it was shown in
figs.~\ref{fig:mpcac_kappa} to~\ref{fig:kappa_zf},
we expected the value of $\kcr$ not to change appreciably even if
$\zfcr$ would vary visibly from one method to another.
Therefore, we considered that the value of $\kcr$ obtained using
(1*) was already a very accurate guess on where the critical value of
$\kappa$ should be using all the other conditions.
In fact, these expectations were later confirmed from our results of $\kcr$,
which, from one method to another, are the same within statistical
errors. Actually, in most cases $\kcr$ did
not change in any digit between any of the methods employed in the
determination of $\zfcr$ (cf. tab.~\ref{tab:kc_beta}).

On the contrary, changes in $\zfcr$
among the different methods are particularly manifest.
This may be seen better in tab.~\ref{tab:zfc_beta}.
Here we can see how, in most cases,
$\zfcr$ does not agree within errors from one method to another.
This behavior becomes stronger at lower energies, i.e.
for decreasing values of $\beta$.
Even if $\zfcr$ does not agree from one method to another,
the differences are expected to be only O($a$) discretization
effects and, as such, should vanish in the continuum limit linearly in the
lattice spacing.

\begin{figure}
\centering
\includegraphics[width=0.7\textwidth]{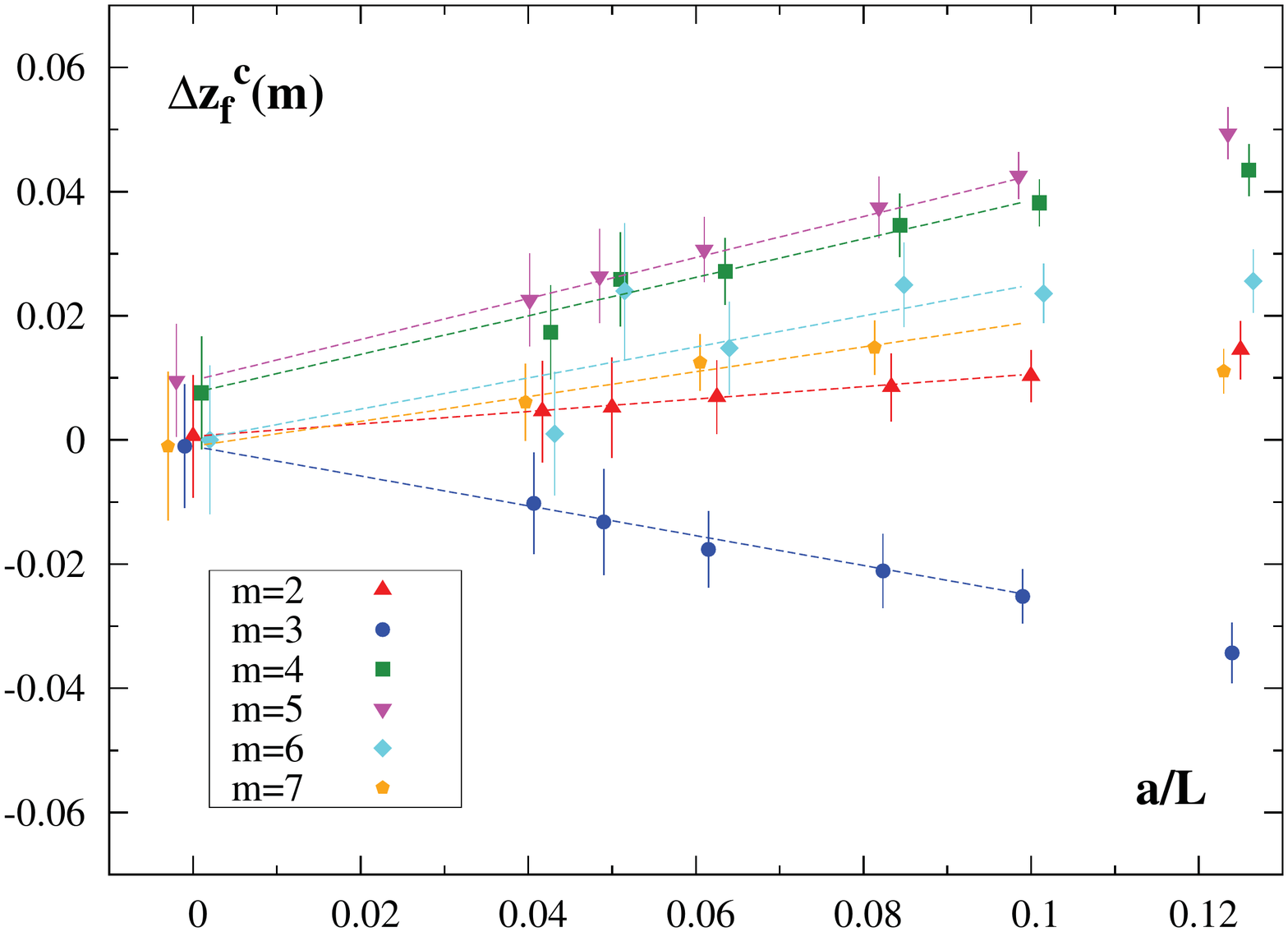}
\caption{Differences of $\zfcr$, $\Delta \zfcr(m)$, as
  determined from different methods (cf. eq.~\eqref{eq:DeltaZf}).
  The differences are always $\zfcr(1)$ minus $\zfcr(m)$, determined
  from any other method $m=2 \dots 7$.
  The data are presented in tab.~\ref{tab:Delta_zf_1M_np_cl}.
  All extrapolations to the continuum limit are linear in $a/L$ and
  the point $L/a=8$ is excluded from all the fits.
  The results from the fits are presented in tab.~\ref{tab:Fit_Delta_zf_1M_np_cl}.
  The points from the different methods have been plotted slightly displaced from each other.}
\label{fig:delta_zf}
\end{figure}

In order to check this expectation, we have performed the continuum limit of
differences in $\zfcr$, as determined from different methods, at
the lowest value of the renormalization scale (cf. also~\cite{Sint:2010xy}).
In particular, the data correspond to the differences
\begin{equation}
\label{eq:DeltaZf}
\Delta \zfcr(m)=\zfcr(1) - \zfcr(m) \, , \quad m=2,3,4,5,6,7 \, .
\end{equation}
The data for $\Delta \zfcr(m)$ are presented in
tab.~\ref{tab:Delta_zf_1M_np_cl} for all the methods and the
corresponding fits, linear in $a/L$, are given in
tab.~\ref{tab:Fit_Delta_zf_1M_np_cl} where the point $L/a=8$ has been excluded from all the fits.
The data for $\Delta \zfcr(m)$, together with the extrapolation to
the continuum limit are plotted in fig.~\ref{fig:delta_zf}.
From this analysis we can conclude that the differences in $\zfcr$
from different methods are only cutoff effects of O($a$), as expected,
which vanish in the continuum limit.
This result may be considered as an additional test of the
universality of the continuum limit. Moreover, discrepancies of O($a$) between different
values of $\zfcr$ should affect physical observables at
$O(a^{2})$ at most.
This expectation will be confirmed in the following section,
where we analyze the dependence of several quantities on the
particular tuning condition.

\subsection{Conclusions on the tuning}
\label{sec:Conclusionsonthetuning}

We have presented the results of the non-perturbative tuning of $\kappa$
and $\zf$ for the $\chi$SF at several physical scales, for a range of
lattice spacings, and using 7 different definitions of $\zfcr$.
Our results demonstrate that the tuning of these two coefficients is indeed
feasible, at least in the quenched approximation.
Moreover, we observe that the tuning of $\zf$ and $\kappa$ are
nearly independent.
This observation is important, having in mind dynamical fermion
simulations; if this behaviour persists with dynamical calculations,
it may ease the numerical effort necessary to perform the tuning,
thus reducing the number of required simulations.
We have also shown that even if $\zfcr$ differs from one method to
another at finite values of the cutoff, such differences are only
O($a$) discretization effects, as expected theoretically.
These discrepancies vanish in the continuum limit,
which itself provides numerical evidence of the universality of the
continuum limit.

\section{Scaling studies and universality of the continuum limit}
\label{sec:scaling}

In the present section,
we show the results of our scaling analysis of several correlation
functions that have been computed using the values of the critical
parameters, $\kcr$ and $\zfcr$,
as determined from each of the 7 tuning conditions defined in
sect.~\ref{ssec:Tuningconditions}.
We have carried out these studies at all the $\beta$
values at which the tuning has been performed.
In sect.~\ref{ssec:Tuningresults} we have shown that
different definitions of $\zfcr$ lead to critical values of
$z_{f}$ that differ from each other by cutoff effects of O($a$).
With the scaling study presented here,
we demonstrate that these discrepancies in $\zfcr$ do not
influence the continuum limit value of physical observables.
This is a very important result since the continuum limit must be
independent of the particular definition of the critical parameters.
Furthermore, we will show that physically relevant quantities,
when determined from the different values of $\zfcr$,
agree within statistical errors already at non-zero
lattice spacing, even at the coarsest lattices.
This agreement holds even at the matching scale with the hadronic
scheme, where cutoff effects are expected to be largest.
Indeed, the agreement at non-zero lattice spacing indicates that the
discretization effects induced by the O($a$) uncertainties in $\zfcr$ are very small.

In order to analyze the different correlation functions,
we have classified them in three types.
The first, discussed in sect.~\ref{ssec:R512evenminus},
are those ${\mathcal R}_5^{1,2}$-even correlation functions
that have a non-vanishing continuum limit, which we refer to as 'physical'. 
These are the only quantities which are expected to be automatic O($a$)-improved (up to boundary effects),
provided $\kappa$ and $z_{f}$ are correctly tuned to their critical values.
The second kind, detailed in sect.~\ref{ssec:R512evenplus},
are those ${\mathcal R}_5^{1,2}$-even
correlation functions which vanish in the continuum limit,
if the correct $\chi$SF b.c.~\eqref{eq:DirichletFermionHomogQcontSpectrum} are recovered 
in the continuum limit.
The last type, sect.~\ref{ssec:R512odd},
are ${\mathcal R}_5^{1,2}$-odd quantities. They should
vanish in the continuum limit up to O($a$) cutoff effects if
${\mathcal R}_5^{1,2}$-symmetry is restored in the continuum limit.

These 3 kinds of correlation functions have been obtained from the
definitions of the boundary to bulk correlation functions given in
eq.~\eqref{eq:bvcfTun}. For unexplained notations in this section, the reader is referred to
sect.~\ref{ssec:somedef}.
With suitable combinations of the Dirac and flavor indices in
eq.~\eqref{eq:bvcfTun}, we can define correlation functions which
are either even or odd under ${\mathcal R}_5^{1,2}$ transformations.
All correlation functions labelled with $+$ should vanish in
the continuum limit if the correct boundary conditions are recovered, independently
of symmetry considerations. On the contrary, all those labelled with $-$
are a priori different from zero, unless some symmetry requires
them to vanish.

Boundary to bulk correlation functions are normalized in a standard fashion with certain
boundary to boundary correlation functions in order to cancel the
renormalization of the boundary quark fields.
In particular in this work, only one such boundary to boundary correlation function
is considered. It is the equivalent of
$f_{1}$~\cite{Luscher:1996iy} in the standard SF and it is defined as,
\begin{equation}
\label{eq:g1Scaling}
g_{1}^{ab}(\theta) = - \frac{1}{L^{6}}\, \langle
\widetilde{\mathcal{P}'}_{+}^{a}\widetilde{\mathcal{P}}_{-}^{b}\rangle
\, .
\end{equation}
Note that the combination of signs in Eq.~\eqref{eq:g1Scaling} is the
only possibility for $g_{1}^{ab}$ not to vanish in the continuum
limit, according to the boundary conditions satisfied in the continuum.

\subsection{${\mathcal R}_5^{1,2}$-even correlation functions}
\label{ssec:R512evenminus}

For 14 values of $\beta$ and several kinematic conditions, we have analyzed 
the ${\mathcal R}_5^{1,2}$-even correlation functions $g_{\mathrm{P}_{-}}^{11}$, $g_{\mathrm{V}_{0 -}}^{12}$ and
$g_{1}^{11}$ as defined in eq.~\eqref{eq:bvcfTun} and in eq.~\eqref{eq:g1Scaling}.
The correlation functions determined at the values of $\kcr$ and $\zfcr$ obtained from
the tuning conditions (1) to (7) for all the $\beta$ values and all the renormalization
scales considered are collected in
tabs.~\ref{tab:ecf_0.0_newtun_np}-\ref{tab:ecf_0.5_newtun_pp}.
In tabs.~\ref{tab:ecf_0.0_newtun_np}-\ref{tab:ecf_0.0_newtun_pp}
we give results at $\boldsymbol{\theta}=(0,0,0)$, while in
tabs.~\ref{tab:ecf_0.5_newtun_np}-\ref{tab:ecf_0.5_newtun_pp}
we collect results for $\boldsymbol{\theta}=(0.5,0.5,0.5)$.

The determination of all these quantities from the methods (1) to (7)
has been performed via interpolations to $\kcr$ and $\zfcr$.
To check that the interpolation does not introduce additional systematic errors,
we have computed observables directly at $\kcr$ and $\zfcr$,
without performing any interpolation from the first tuning method 
(cf. method (1*) in tabs.~\ref{tab:kc_beta} and ~\ref{tab:zfc_beta}).
These data, computed at $\boldsymbol{\theta}=(0.5,0.5,0.5)$, are
presented in tabs.~\ref{tab:ecf_0.5_newtun_np}-\ref{tab:ecf_0.5_newtun_p}.
We have checked for this particular tuning method that the quantities obtained 
via an interpolation of the data
do agree within errors with those obtained by means of
new computations performed directly at the critical values of $\kappa$ and $z_{f}$, denoted as method (1*).
We therefore believe that the interpolation that we perform here does not induce additional errors.

\begin{figure}
  \centering
  \includegraphics[width=0.70\textwidth,angle=270]{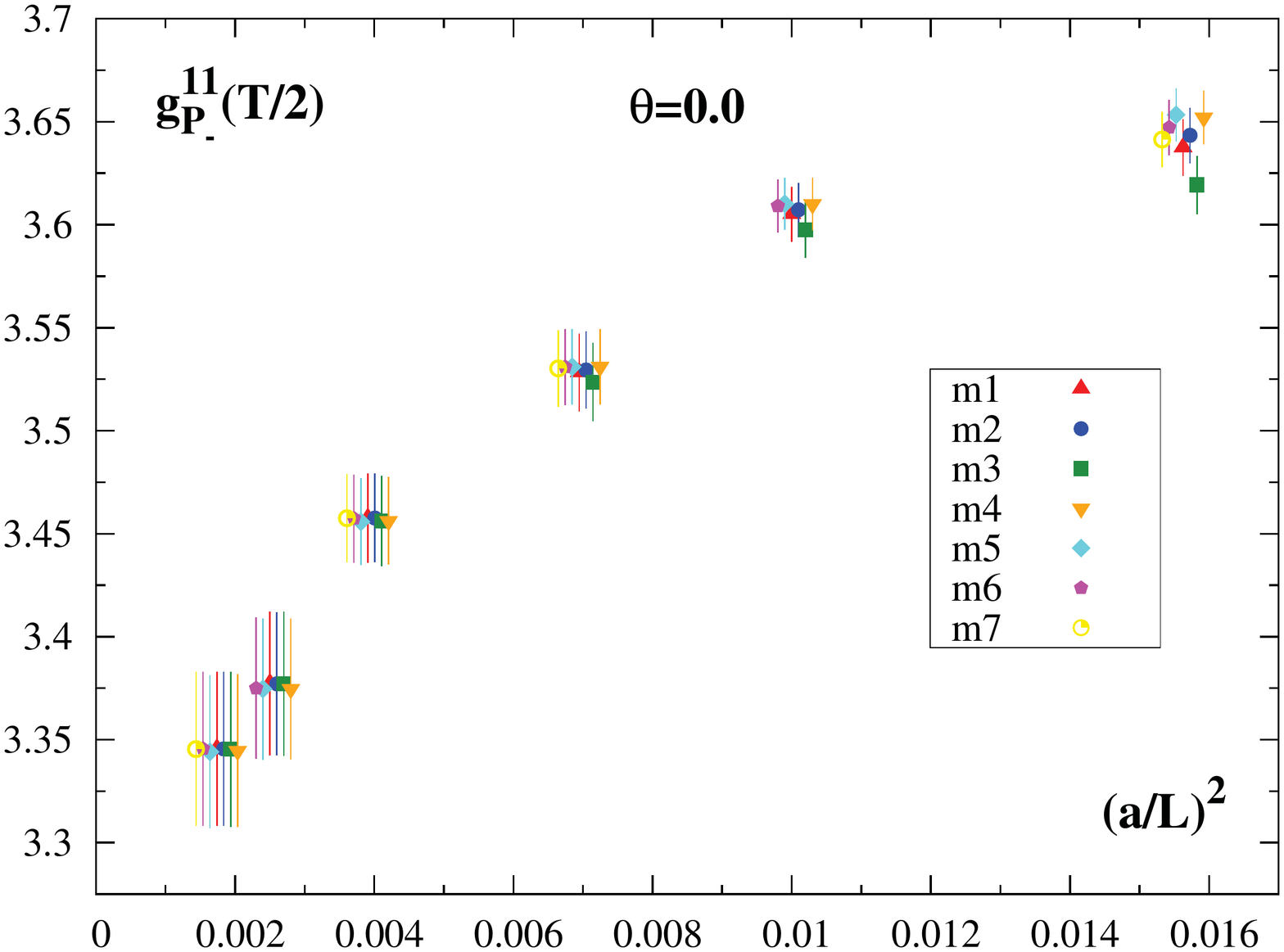}
  \caption{Comparison between different tuning conditions using the
    ${\mathcal R}_5^{1,2}$-even quantity $g_{\mathrm{P}_{-}}^{11}$.
    The scale is NP and $\boldsymbol{\theta}=(0,0,0)$.
    Data for all methods (1) to (7) are presented
    (cf. tab.~\ref{tab:ecf_0.0_newtun_np}).
    No continuum limit is performed, since this quantity takes a
    finite value in the continuum limit only after renormalization.
    The purpose of the plot is to compare the results from
    the different tuning conditions at non-zero lattice spacing and a full
    agreement can be seen.
    The data from the different methods have been plotted slightly
    displaced from each other.}
  \label{fig:gpm_cl_np_0.0}
\end{figure}

From this analysis we have found that 
for all the 14 values of $\beta$, all ${\mathcal R}_5^{1,2}$-even quantities with a non-vanishing
continuum limit do not depend on
the definition of $\zfcr$ within statistical errors.
This holds for any of the values of the kinematical parameters that we
have investigated.
In fig.~\ref{fig:gpm_cl_np_0.0} we show, as an example, the dependence on the tuning
conditions for $g_{\mathrm{P}_{-}}^{11}$ at the matching scale, $L=1.436\,r_{0}$, and for 
$\boldsymbol{\theta}=(0,0,0)$. Similar results are obtained for the other correlation functions.

The independence at each value of the lattice spacing on the tuning conditions, 
i.e. on the particular definition of $\zfcr$ adopted is reassuring: no large cutoff effects are introduced 
depending on the choice of the tuning condition.
The continuum limit of renormalized quantities and their dependence on the tuning conditions
will be discussed in the companion paper~\cite{Lopez:2012mc}.

\subsection{Recovery of the $\chi$SF boundary conditions}
\label{ssec:R512evenplus}

We present here results for the ${\mathcal R}_5^{1,2}$-even correlation
functions which should vanish in the continuum limit, due to the particular form of the 
$\chi$SF boundary conditions in eqs.~(\ref{eq:DirichletFermionHomogQcontSpectrum},
\ref{eq:DirichletAntiFermionHomogQcontSpectrum}) satisfied by the fermion fields in the continuum limit.
We collect our results for the $g_{\mathrm{P}_{+}}^{11}$ and $g_{\mathrm{V}_{0 +}}^{12}$ correlation functions 
at all the $\beta$ values and all the renormalization conditions in
tabs.~\ref{tab:ecf_0.0_newtun_np}-\ref{tab:ecf_0.5_newtun_pp}.
We provide results for $\boldsymbol{\theta}=(0,0,0)$ in 
tabs.~\ref{tab:ecf_0.0_newtun_np}-\ref{tab:ecf_0.0_newtun_pp}
and for $\boldsymbol{\theta}=(0.5,0.5,0.5)$ in
tabs.~\ref{tab:ecf_0.5_newtun_np}-\ref{tab:ecf_0.5_newtun_pp}.

We found that for all the tuning conditions $g_{\mathrm{P}_{+}}^{11}$ and $g_{\mathrm{V}_{0 +}}^{12}$
vanish in the continuum limit, signalling that the proper $\chi$SF b.c. are recovered.
We have observed that the values of $g_{\mathrm{P}_{+}}^{11}$ and $g_{\mathrm{V}_{0 +}}^{12}$
would change substantially if we vary the values of $\zfcr$ within their statistical
errors. If we want to perform the continuum limit at fixed ``physical'' scale
we have to take this variation into account\footnote{We acknowledge a very important discussion with S.Sint
on this point.}. We do this by propagating
the statistical error of $\zfcr$ when interpolating $g_{\mathrm{P}_{+}}^{11}$ and 
$g_{\mathrm{V}_{0 +}}^{12}$ in $\zf$. We note that $g_{\mathrm{P}_{+}}^{11}$ and $g_{\mathrm{V}_{0 +}}^{12}$
do not show the same behaviour for variations of $\kcr$.
 
\begin{figure}
  \centering
  \includegraphics[width=0.8\textwidth]{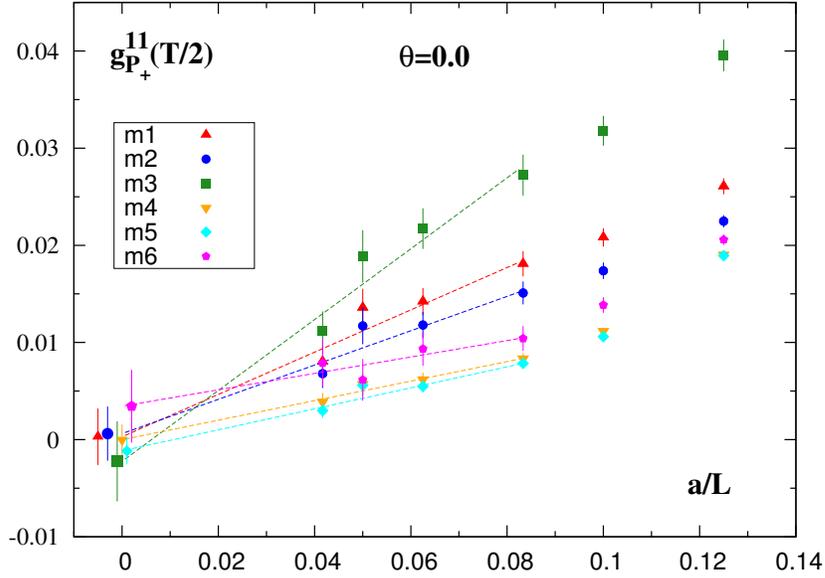}
  \caption{Approach to the continuum limit of the
    ${\mathcal R}_5^{1,2}$-even quantity $g_{\mathrm{P}_{+}}^{11}$.
    The scale is NP and $\boldsymbol{\theta}=(0,0,0)$.
    Data for methods (1) to (6) are presented
    (cf. tab.~\ref{tab:ecf_0.0_newtun_np}).
    $g_{\mathrm{P}_{+}}^{11}$ is plotted here as a function of $a/L$.
    The error bars contain the statistical errors of the correlation
    functions together with the propagation of the statistical error on $\zfcr$.}
  \label{fig:gpp_cl_np_0.0}
\end{figure}

We show in fig.~\ref{fig:gpp_cl_np_0.0} the behaviour of 
$g_{\mathrm{P}_{+}}^{11}(\boldsymbol{\theta=\boldsymbol{0}})$ towards the continuum limit.
This is an example
of a quantity that should vanish in the continuum limit if the proper
b.c. conditions are recovered. Similar plots can be obtained for other
quantities looking at the data in tabs.~\ref{tab:ecf_0.0_newtun_np}-\ref{tab:ecf_0.5_newtun_pp}
where the first error is statistical while the second, where available, contains the propagation
of the error of $\zfcr$.
As discussed in the previous and in the next section, we stress that this is the only case where we have 
observed that the statistical error of $\zfcr$ shows significant effects
in the correlation functions.

\subsection{${\mathcal R}_5^{1,2}$-odd correlation functions}
\label{ssec:R512odd}

Amongst all the ${\mathcal R}_5^{1,2}$-odd correlation functions, 
we consider correlation functions that vanish only because of symmetry considerations, 
i.e. we do not consider correlation functions vanishing
in the continuum limit because of the recovery of the $\chi$SF boundary conditions.
Examples of such correlation functions are
$g_{\mathrm{A}_{0 -}}^{11}$ and $\overline{g}_{\mathrm{V}_{k -}}^{12}$.
It is important to note that the same correlation functions have been used in the determination
of $\zfcr$.
When studying the dependence on the tuning condition towards the continuum limit of a given 
correlation function we have obviously excluded the value
of $\zfcr$ obtained imposing that the same correlation function vanishes at each value of the lattice 
spacing. We have though studied the dependence of all the other tuning conditions 
towards the continuum limit.

The numerical results obtained at $\boldsymbol{\theta}=(0,0,0)$ are presented in
tabs.~\ref{tab:ocf_0.0_newtun_np}-\ref{tab:ocf_0.0_newtun_pp},
and the results for $\boldsymbol{\theta}=(0.5,0.5,0.5)$
are given in tabs.~\ref{tab:ocf_0.5_newtun_np}-\ref{tab:ocf_0.5_newtun_pp}.

Although results from different definitions of $\zfcr$ do not
coincide at non-zero lattice spacing, all the ${\mathcal R}_5^{1,2}$-odd
correlation functions vanish in the continuum limit independently of the 
tuning condition adopted. 
This is strong evidence that ${\mathcal R}_5^{1,2}$-symmetry is restored in
the continuum limit.

\begin{figure}
\centering
  \includegraphics[width=0.9\textwidth]{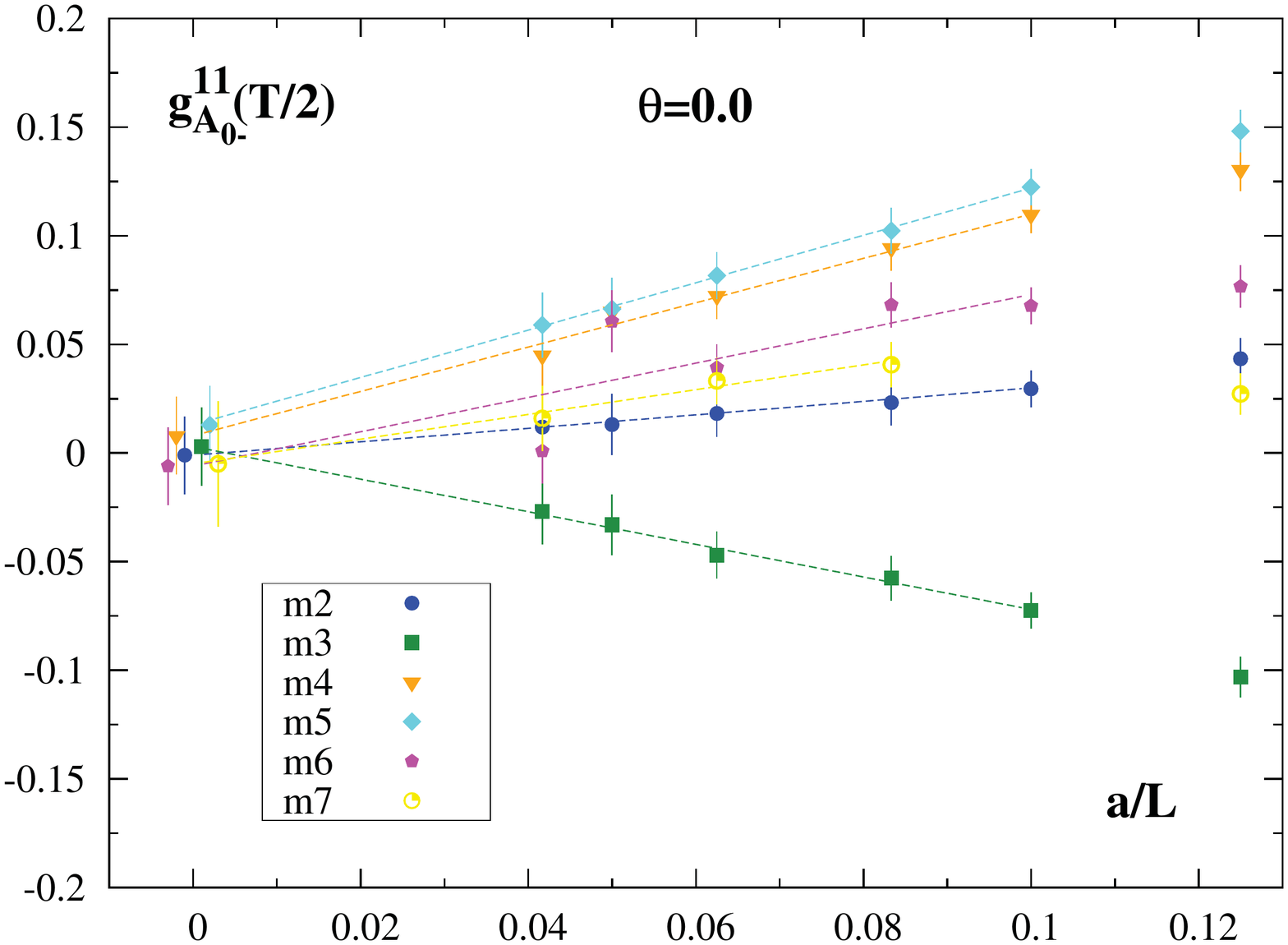}
  \caption{Extrapolation to the continuum limit of the
    ${\mathcal R}_5^{1,2}$-odd quantity $g_{\mathrm{A}_{0 -}}^{11}$.
    The scale is NP and $\boldsymbol{\theta}=(0,0,0)$.
    Data for  methods (2) to (7) are presented
    (cf. tab.~\ref{tab:ocf_0.0_newtun_np}).
    The fits are all linear in $a/L$
    (cf. tab.~\ref{tab:ga0_cl_np_fit}).
    The point $L/a=8$ has been excluded from all the fits.
    We show the data for all tuning conditions except condition (1)
    because it corresponds to imposing $g_{\mathrm{A}_{0 -}}^{11}=0$.
    The (vanishing) continuum limit values obtained from the different tuning
    conditions have been plotted slightly displaced from each other.}
  \label{fig:ga0_cl_np_0.0}
\end{figure}

\begin{figure}
  \includegraphics[width=0.63\textwidth,angle=270]{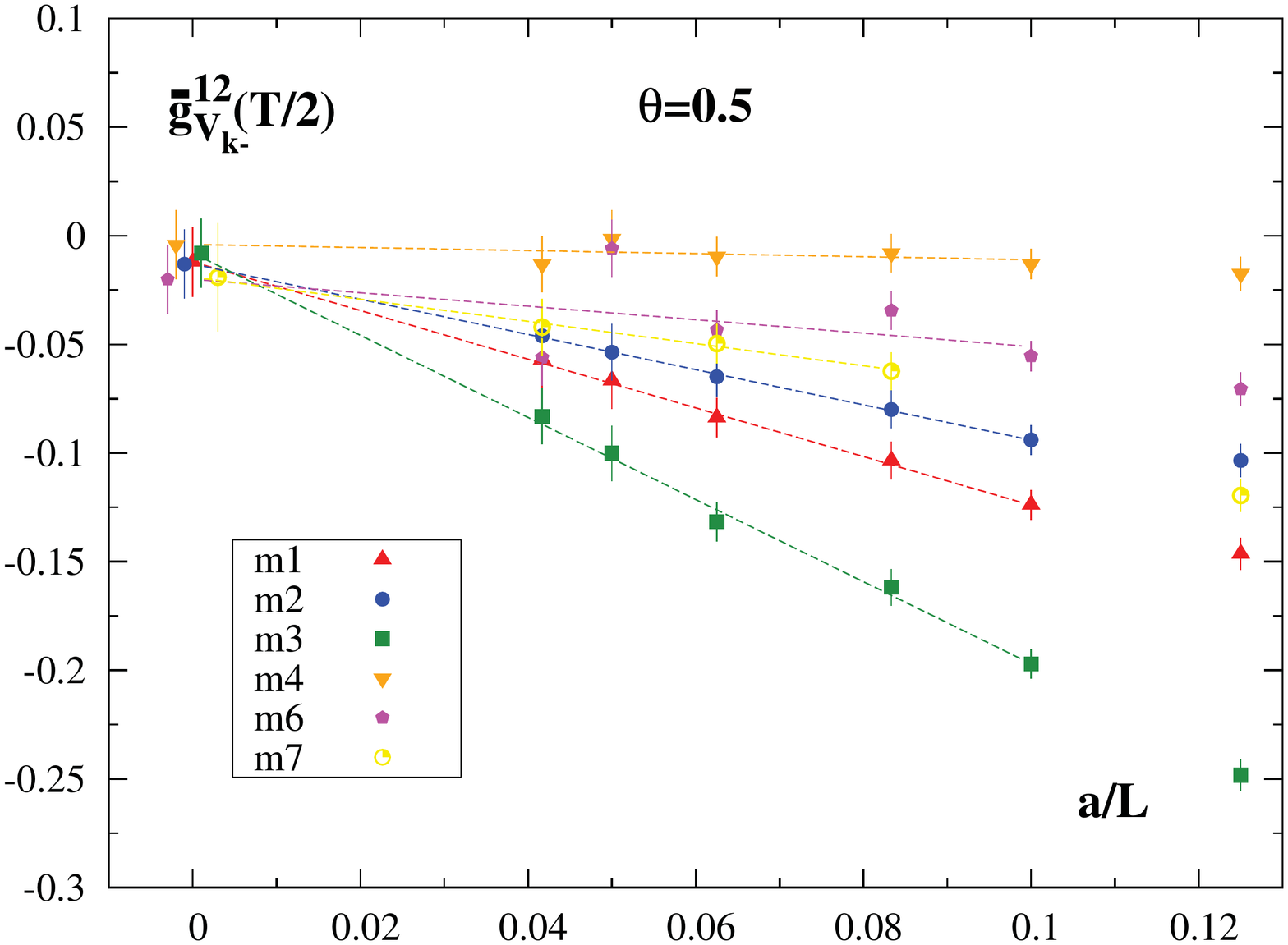}
\caption{Extrapolation to the continuum limit of the
    ${\mathcal R}_5^{1,2}$-odd quantity $\overline{g}_{\mathrm{V}_{k -}}^{12}$.
    The scale is NP and $\boldsymbol{\theta}=(0.5,0.5,0.5)$.
    Data for  methods (1) to (4) and (6) to (7) are presented
    (cf. tab.~\ref{tab:ocf_0.5_newtun_np}).
    The fits are all linear in $a/L$
    (cf. tab.~\ref{tab:gak_cl_np_fit}).
    The point $L/a=8$ has been excluded from all the fits.
    We show the data for all tuning conditions except condition (5)
    because it corresponds to imposing $\overline{g}_{\mathrm{V}_{k -}}^{12}=0$.}
  \label{fig:gak_cl_np_0.5}
\end{figure}

As an example we show in the plot of fig.~\ref{fig:ga0_cl_np_0.0} 
the continuum limit approach for $g_{\mathrm{A}_{0 -}}^{11}$ with $\boldsymbol{\theta}=(0,0,0)$ 
and in the plot of fig.~\ref{fig:gak_cl_np_0.5} the continuum limit of 
$\overline{g}_{\mathrm{V}_{k -}}^{12}$ for $\boldsymbol{\theta}=(0.5,0.5,0.5)$.
These plots correspond to our most non-perturbative point,
$L=1.436\,r_{0}$, which is the case where the cutoff effects are expected to be
stronger. The data show a linear behaviour in $a/L$ and they have been fitted with a 
linear fit in $a/L$,
\begin{equation}
\label{eq:FitBbisbis}
f = b_{0} + b_{1}\Big(\frac{a}{L}\Big) \, .
\end{equation}
We have not considered the point $L/a=8$ in the fits.
The results from the fits are summarized in
tab.~\ref{tab:ga0_cl_np_fit} for $g_{\mathrm{A}_{0 -}}^{11}$
and in tab.~\ref{tab:gak_cl_np_fit}
for $\overline{g}_{\mathrm{V}_{k -}}^{12}$.
As anticipated all the ${\mathcal R}_5^{1,2}$-odd correlation functions vanish 
within errors in the continuum limit. Not being automatic O($a$) improved they scale 
linearly in $a/L$ independently of the tuning condition adopted.

\subsection{Conclusions on the scaling analysis}
\label{ssec:Conclusionsonthescalinganalysis}

From the analysis performed at several $\beta$ values at fixed renormalization scale, 
we learned that all quantities that should vanish in the continuum limit,
either by boundary conditions or symmetry restoration,
have a visible dependence on the tuning condition and the corresponding values
of $\zfcr$.

The ${\mathcal R}_5^{1,2}$-even correlation functions that should vanish in the 
continuum limit, if the correct boundary conditions are recovered, show the strongest dependence
on the way $\zfcr$ has been determined. 
In particular we have observed that to properly study the continuum limit
we cannot simply take the central vale of $\zfcr$, but we need to propagate the statistical error
of $\zfcr$ into the error for the correlation functions, i.e. the statistical fluctuations in $\zfcr$ are 
visible in the final values for the correlation functions. 
If one takes the statistical error of $\zfcr$ into account in the determination of the error
of the correlation functions, we observe, for all the tuning conditions, that the proper boundary
conditions are recovered in the continuum limit within errors. Additionally this result
is confirmed for all the tuning conditions adopted.

This observation is in stark contrast with the ${\mathcal R}_5^{1,2}$-even correlation functions
containing the non-vanishing component of the boundary fermion fields. 
In this case not only the statistical error
on $\zfcr$ is irrelevant, but even the choice of the tuning condition for all practical purposes 
does not change the final values for the correlation functions even at non-zero lattice spacings.
Although we can not make a general statement about all possible physical observables,
our expectation is that other physical quantities,
different from the ones studied in this work, will behave in the same manner.
This is reassuring for further uses of the $\chi$SF for physical applications with dynamical 
fermions. 

In the case of the ${\mathcal R}_5^{1,2}$-odd orrelation functions,
we have shown that they vanish in the
continuum limit with leading O($a$) discretization effects for all the
tuning conditions, as expected.
This is strong numerical evidence of the restoration of
${\mathcal R}_5^{1,2}$-symmetry in the continuum limit and the
independence of the continuum limit on the particular tuning
condition.
This result is also an indication that the correct boundary
conditions are recovered in the continuum limit.

\section{Conclusions}
\label{sec:conclu}

Large scale simulations with Wilson twisted mass fermions at maximal twist 
need a renormalization scheme that preserves the property of automatic O($a$) improvement.
The RI-MOM scheme is consistent with this requirement and results for $N_f=2$ dynamical fermions
have been obtained in ref.~\cite{Constantinou:2010gr} and preliminary
$N_f=4$ results have been presented in ref.~\cite{ETM:2011aa}. 
Recently the x-space scheme has been tested on the $N_f=2$ twisted mass ensembles~\cite{Cichy:2012is}.
One of the problems with these schemes is that it is difficult to cover the large range 
of scales necessary to bridge the perturbative and non-perturbative regimes.

Finite volume schemes have been developed to tackle this problem. These schemes can be used
to perform the continuum limit of step-scaling functions and to carry out 
the non-perturbative renormalization, especially for scale-dependent quantities.
Thus it is very desirable to have the possibility to use finite volume renormalization schemes 
together with Wilson twisted mass fermions. In this work we have made a detailed study of the $\chi$SF scheme
proposed in ref.~\cite{Sint:2010eh} with quenched Wilson fermions. 
Bulk automatic O($a$) improvement is achieved with a single non-perturbative tuning of the parameter $z_f$ to its critical
value $\zfcr$,
together with the usual tuning of $\kcr$.
We have proposed a tuning strategy for the 
simultaneous determination of $\zfcr$ and $\kcr$ and we have perfomed a feasibility
study of several tuning conditions, showing that the tuning is affordable with
the current computer resources and it does not pose any specific problem.

The study presented in this work has shown that, as expected, unphysical correlation functions
vanish in the continuum limit with O($a$) corrections independently of the tuning condition
adopted and that the proper boundary conditions are recovered in the
continuum limit. 
In addition physical correlation functions are rather insensitive to the method employed to determine
$\kcr$ and $\zfcr$. This is a very promising result 
since it will ease the computational effort required to go
beyond the quenched approximation.

In a companion paper~\cite{Lopez:2012mc} we will further investigate the continuum limit
scaling behaviour of physical quantities that are renormalized through the $\chi$SF scheme
employing the results of $\kcr$ and $\zfcr$ determined in this work.

\section*{Acknowledgments}
\vspace{-1.0cm}

We thank S. Sint and B. Leder for many useful discussions. 
We also acknowledge the support of the computer center in DESY-Zeuthen and the NW-grid in Lancaster.
This work has been supported in part by the DFG Sonderforschungsbereich/Transregio SFB/TR9-03.
This manuscript has been coauthored by Jefferson Science Associates, LLC under Contract 
No. DE-AC05-06OR23177 with the U.S. Department of Energy.

\newpage

\begin{appendix}

\section{Symmetries in the twisted basis}
\label{app:CPTtwisted}

We list in this appendix the form of a few symmetry tranfosrmations in the twisted basis.
The twisted basis $\{ \chi \, , \overline{\chi}\}$ 
and the standard basis $\{ \psi \, , \overline{\psi}\}$
are related by the non-anomalous axial transformation
\begin{equation}
\label{eq:RotationStoTw}
\psi(x) = \exp\left(i \, \frac{\omega}{2} \, \gamma_{5} \, \tau^{3} \right) \, \chi(x) \, ,
\qquad
\overline{\psi}(x) = \overline{\chi}(x) \, \exp\left(i \,
  \frac{\omega}{2} \, \gamma_{5} \, \tau^{3} \right) \, .
\end{equation}

Charge conjugation $\mathcal{C}$ is invariant under this basis transformation
\be
\mathcal{C} \colon
\begin{cases}
   U(x;\mu) \rightarrow  U(x;\mu)^*, \\
   \chi(x)    \rightarrow  
                C^{-1}\chibar(x)^T,\\
   \chibar(x) \rightarrow 
               - \chi(x)^T C ,
\end{cases}
\label{eq:chargeconj}
\ee
where $C$ satisfies
\be
- \gamma_\mu^T = C \gamma_\mu  C^{-1} , \qquad \gamma_5 = C \gamma_5  C^{-1}. 
\ee
Parity and time reversal are affected by the rotation and take a
different form in the twisted basis.
In this basis they are denoted $\mathcal{P}_{\omega}$ and
$\mathcal{T}_{\omega}$, respectively, and have the expressions
\begin{alignat}{2}
\label{eq:Ptw}
\mathcal{P}_{\omega} \, : \,
\begin{cases}
U_{0}(x_{0},\vec{x}) \, \rightarrow \, U_{0}(x_{0}, -\vec{x}) \, ,
\quad
U_{k}(x_{0},\vec{x}) \, \rightarrow \, U_{k}^{\dagger}(x_{0}, -\vec{x} -
\vec{k}) \\
\chi(x_{0},\vec{x}) \, \rightarrow \, \gamma_{0} \, \exp\left( i \,
  \omega \, \gamma_{5} \, \tau^{3} \right) \, \chi(x_{0}, -\vec{x})\\
\overline{\chi}(x_{0}, \vec{x}) \, \rightarrow \, \overline{\chi}(x_{0}, -\vec{x})\, 
\exp\left( i \, \omega \, \gamma_{5} \, \tau^{3} \right) \,\gamma_{0}
\end{cases}
\end{alignat}

\begin{alignat}{2}
\label{eq:Ttw}
\mathcal{T}_{\omega} \, : \,
\begin{cases}
U_{0}(x_{0},\vec{x}) \, \rightarrow \, U_{0}^{\dagger}(-x_{0} - a, \vec{x}) \, ,
\quad
U_{k}(x_{0},\vec{x}) \, \rightarrow \, U_{k}(-x_{0}, \vec{x}) \\
\chi(x_{0},\vec{x}) \, \rightarrow \, i\, \gamma_{0} \,
\gamma_{5} \, \exp\left(i \,  \omega \, \gamma_{5} \, \tau^{3} \right)
\,\chi(-x_{0}, \vec{x})\\ 
\overline{\chi}(x_{0}, \vec{x}) \, \rightarrow \, \overline{\chi}(-x_{0},
\vec{x})\, \exp\left( i \,
  \omega \, \gamma_{5} \, \tau^{3} \right) \, i \, \gamma_{0} \, \gamma_{5}\,.
\end{cases}
\end{alignat}
To obtain the form of a parity or a time-reversal transformation in the standard basis, $\mathcal{P}$ or 
$\mathcal{T}$, it is sufficient to set $\omega=0$ in eqs~\eqref{eq:Ptw} and~\eqref{eq:Ttw}.

In the twisted basis the $SU(2)$ vector transformation takes the form
\be
SU(2)_\omega \colon
\begin{cases}
   \chi(x)     \rightarrow 
               \exp\left(-i\, \frac{\omega}{2}\, \gamma_5\, \tau^3\right)\,
\exp\left(i\, \frac{\alpha_V^a}{2}\, \tau^a\right)\,
\exp\left(i\, \frac{\omega}{2}\,\gamma_5\,\tau^3\right)\,\chi(x),\\
   \chibar(x)  \rightarrow 
               \chibar(x)\,\exp\left(i\,\frac{\omega}{2}\,\gamma_5\,\tau^3\right)\,
\exp\left(-i\,\frac{\alpha_V^a}{2}\,\tau^a\right)\,
\exp\left(-i\,\frac{\omega}{2}\,\gamma_5\,\tau^3\right) .
\end{cases}
\label{eq:tv}
\ee

\newpage
\section{The free lattice quark propagator for the $\chi$SF with Wilson fermions}
\label{app:freeWchiSF}

In this appendix we obtain the analytical expression of the quark propagator $S\left( x,y \right)$
on the lattice at tree-level of perturbation theory for the action 
in eq.~\eqref{eq:FactionWChiSF}.
The derivation of the propagator is rather standard
(see for example ref.~\cite{Luscher:1996vw}) once we have the exact b.c. satisfied by the fermion
fields at finite lattice spacing $a$. They can be obtained as a spinoff of the 
orbifold construnctions~\cite{Sint:2010eh} and read
\be
\widetilde{Q}_{+}\left(1 - \frac{a}{2}\partial^{*}_{0} \right)S(x,y)|_{x_{0} = 0} =0
\qquad
\widetilde{Q}_{-}\left(1 + \frac{a}{2}\partial_{0} \right)S(x,y)|_{x_{0} = T} =0 \, .
\label{eq:WchiSF_bc}
\ee
The problem we want to solve is then
\be
\left(D_{\textrm{W}} + m_{0} \right)\,S\left( x,y \right) = a^{-4}\,
\ee
with b.c.~\eqref{eq:WchiSF_bc} where $\delta_{x,y}$ is the dimensionless Kronecker delta
and $D_{W}$ denotes the massless Wilson operator~\eqref{eq:WDopInTermsOfK}.

The result of the calculation can be cast in the form
\begin{equation}
\label{eq:QuarkPropFinalTimeMomLat}
S \left( x,y \right) = \left( D_{\textrm{W}}^{\dagger} + m_{0}
\right)\, G\left( x,y \right) \, ,
\end{equation}
with
\begin{equation}
\label{eq:QuarkPropFinalLat}
G\left( x,y \right) = \frac{1}{L^{3}}\, \sum_{\vec{p}}\,
e^{i\vec{p}\left(\vec{x}-\vec{y}\right)}\,
G\left( x_{0},y_{0};\vec{p}\right) \, ,
\end{equation}
where
\begin{equation}
\label{eq:GPropFinalLatPhase}
\begin{split}
G\left( x_{0},y_{0};\vec{p} \right) = \frac{1}{2 \opall(\vec{p}^{+}) A(\vec{p}^{+})D(p^{+})}& \Big\{
\, \: e^{-\omega(\vec{p}^{+})|x_{0} - y_{0}|} - e^{-\omega(\vec{p}^{+})\left( 2\left( T + a \right) - |x_{0} - y_{0}|\right)}\\
& 
- i\gamma_{0}\gamma_{5}\tau^{3}e^{-\omega(\vec{p}^{+})\left(x_{0} + y_{0} + a \right)}\\
&+ i\gamma_{0}\gamma_{5}\tau^{3}e^{-\omega(\vec{p}^{+})\left( 2\left(T
      + a\right) - \left(x_{0} + y_{0} + a \right) \right)}
\: \Big\} \, .
\end{split}
\end{equation}
Here we list all the definitions of the functions useful for the determination of the propagator.

\begin{equation}
\label{eq:MomPhaseDef}
p_{\mu}^{\pm} = p_{\mu} \pm \theta_{\mu}/L \, , \qquad \theta_{0}=0 \, ,
\end{equation}
\begin{equation}
\label{eq:MomAndMassLatPhase}
\ppall_{\mu}^{\pm} = \frac{1}{a}\sin\left( ap_{\mu}^{\pm}\right) \, ,
\qquad
\mathcal{M}(p^{\pm}) = m_{0} + \frac{1}{2}a\hat{p}^{\pm \, 2}_{\mu}\, ,
\qquad
\hat{p}_{\mu}^{\pm} = \frac{2}{a}\sin\left(
  \frac{ap_{\mu}^{\pm}}{2}\right) \, .
\end{equation}

The function $\omega(\vec{p}^{+})$, such that $p_{0} =
i\omega(\vec{p}^{+})$, is given by
\begin{equation}
\label{eq:DefOmegaLatPhase}
\sinh \left[ \frac{a}{2}\omega \left( \vec{p}^{\pm} \right) \right] = 
\left\{ \frac{a^{2}\ppall ^{\pm \, 2}_{k} + \left( A(\vec{p}^{\pm} ) -1 \right) ^{2}}
{4A(\vec{p}^{\pm})} \right\} ^{1/2} \, , 
\end{equation}
and
\begin{equation}
\label{eq:OmegaPallDefPhase}
\opall(p^{\pm}) \equiv -i\ppall_{0}^{\pm} = \frac{1}{a}\, \sinh\left[
  a\omega(\vec{p}^{\pm}) \right] \, ,
\end{equation}
\begin{equation}
\label{eq:DefALatPhase}
A\left( \vec{p}^{\pm} \right) \equiv 1 + a\left( m_{0} +
  \frac{a}{2}\hat{p}_{k}^{\pm \, 2} \right) \, ,
\end{equation}
\begin{equation}
\label{eq:DFactorDefPhase}
D(p^{\pm}) \equiv 1 + e^{-2\omega(\vec{p}^{\pm})\left(T + a \right)} \, .
\end{equation}

The analytical expression of the lattice quark propagator~\eqref{eq:QuarkPropFinalTimeMomLat}
has been numerically cross-checked with the propagator obtained from a numerical inversion of the free
lattice Dirac operator given in Eq.~\eqref{eq:WDopChiSF}
and also with the corresponding propagator in~\cite{Sint:2010eh}.

\newpage
\section{Tables of numerical results}
\label{app:appC}

\subsection{Tuning data}
\label{app:Notesonthetuning}
\scriptsize

\begin{longtable}[c]{|r|l||r|l|l||r|l|l|}
\hline
\multicolumn{8}{|c|}{Guess values for the tuning} \\
\multicolumn{8}{|c|}{Hadronic scale: $L=1.436\, r_{0}$} \\
\hline
\multicolumn{1}{|c|}{$L/a$} &
\multicolumn{1}{|c||}{$\beta$} &
\multicolumn{1}{|c|}{$N_{\textrm{conf}}$} &
\multicolumn{1}{|c|}{$z_{f}$ } &
\multicolumn{1}{|c||}{$\kappa$ } &
\multicolumn{1}{|c|}{$N_{\textrm{conf}}$} &
\multicolumn{1}{|c|}{$z_{f}$ } &
\multicolumn{1}{|c|}{$\kappa$ } \\
\hline
  8 & 6.0219 & 1000 & 1.74 &  0.1534 & 1000 & 1.79 & 0.1530\\
     &             &          & 1.77 &  0.1537 &          & 1.80 & 0.1534\\
     &             &          & 1.80 &              &          & 1.81 & 0.1537 \\
     &             &          & 1.83 &              &          & 1.82 & 0.1540 \\
     &             &          & 1.86 &              &          & &\\
\hline
10 & 6.1628 & 1000 & 1.73 & 0.1521 & 1000 & 1.78 & 0.1520\\
     &             &          & 1.76 & 0.1522 &      & 1.79 & 0.1521 \\
     &             &          & 1.79 &  &      & 1.80 & 0.1522\\
     &             &          & 1.82 &  &      & 1.81 & 0.1523\\
\hline
12 & 6.2885 &   500 & 1.71 & 0.15050 & 300  & 1.70 & 0.15025\\
     &             &          & 1.74 & 0.15100 &      & 1.73 & 0.15050 \\
     &             &          & 1.77 &  &      & 1.77 & 0.15100 \\
     &             &          & 1.80 &  &      & 1.80 & 0.15125 \\
\hline
16 & 6.4956 &   300 & 1.64 & 0.1489 & 100  & 1.70 & 0.1488\\
     &             &          & 1.67 & 0.1490 & &1.71 & 0.1489\\
     &             &          & 1.70 &  && 1.73 & 0.1490\\
     &             &          & 1.73 &  && 1.74 & 0.1491 \\
     &             &          & 1.76 &  &&&\\
\hline
20 & 6.6790 &   112 & 1.66 & 0.1473 & & &\\
     &             &          & 1.68 & 0.1474 & & &\\
     &             &          & 1.70 & 0.1475 & & &\\
     &             &          & 1.72 & 0.1476 & & &\\
\hline
24 & 6.8187 &   100 & 1.60 & 0.1463 & & &\\
     &             &          & 1.63 & 0.1464 & & &\\
     &             &          & 1.66 & 0.1465 & & &\\
     &             &          & 1.69 & 0.1466 & & &\\
\hline
\caption{Values of $\kappa$ and $z_{f}$ used for the tuning
  and number of configurations, $N_{\textrm{conf}}$, used in all
  calculations at the corresponding value of $\beta$. Scale NP (see text).
  The data of the last three columns have been used only for a
  separate analysis with method (1), which we denote here as method (1*).}
\label{tab:tuning_guess_np_new}
\end{longtable}
\newpage
\begin{longtable}[c]{|r|l||r|l|l||r|l|l|}
\hline
\multicolumn{8}{|c|}{Guess values for the tuning} \\
\multicolumn{8}{|c|}{Intermediate scale: $\overline{g}^{2}=2.4484$}\\
\hline
\multicolumn{1}{|c|}{$L/a$} &
\multicolumn{1}{|c||}{$\beta$} &
\multicolumn{1}{|c|}{$N_{\textrm{conf}}$} &
\multicolumn{1}{|c|}{$z_{f}$ } &
\multicolumn{1}{|c||}{$\kappa$ } &
\multicolumn{1}{|c|}{$N_{\textrm{conf}}$} &
\multicolumn{1}{|c|}{$z_{f}$ } &
\multicolumn{1}{|c|}{$\kappa$ } \\
\hline
  8 & 7.0197 & 1000 & 1.51 & 0.14445 & 1000 & 1.35 & 0.14440\\
     &             &          & 1.54 & 0.14450 &          & 1.45 & 0.14445 \\
     &             &          & 1.57 &               &          & 1.55 & 0.14450\\
     &             &          & 1.60 &               &          & 1.65 & 0.14455\\
\hline
12 & 7.3551 &   500 & 1.46 & 0.1431 &  300 & 1.50 & 0.1430\\
     &             &          & 1.49 & 0.1432 &         & 1.51 & 0.1431\\
     &             &          & 1.52 &             &         & 1.52 & 0.1432\\
     &             &          & 1.55 &             &         & 1.53 & 0.1433\\
\hline
16 & 7.6101 &   300 & 1.44 & 0.1421 &  100 & 1.48 & 0.1420\\
     &             &          & 1.47 & 0.1422 &         & 1.49 & 0.1421\\
     &             &          & 1.50 &             &         & 1.50 & 0.1422\\
     &             &          & 1.53 &             &         & 1.51 & 0.1423\\
\hline
\caption{ Same caption as in tab.~\ref{tab:tuning_guess_np_new} but at
  scale I (see text).}
\label{tab:tuning_guess_int_new}
\end{longtable}

\begin{longtable}[c]{|r|l||r|l|l||r|l|l|}
\hline
\multicolumn{8}{|c|}{Guess values for the tuning} \\
\multicolumn{8}{|c|}{Perturbative scale: $\overline{g}^{2}=0.9944$}\\
\hline
\multicolumn{1}{|c|}{$L/a$} &
\multicolumn{1}{|c||}{$\beta$} &
\multicolumn{1}{|c|}{$N_{\textrm{conf}}$} &
\multicolumn{1}{|c|}{$z_{f}$ } &
\multicolumn{1}{|c||}{$\kappa$ } &
\multicolumn{1}{|c|}{$N_{\textrm{conf}}$} &
\multicolumn{1}{|c|}{$z_{f}$ } &
\multicolumn{1}{|c|}{$\kappa$ } \\
\hline
 8 & 10.3000 & & & & 1000 & 1.2955 & 0.13541 \\
   &         &     & & & & 1.2965 & 0.13544 \\
   &         &     & & & & 1.2975 & 0.13547 \\
   &         &     & & & & 1.2985 & 0.13550 \\
\hline
12 & 10.6086 & & & &  300 & 1.292 & 0.13514 \\
   &         &      & & & & 1.294 & 0.13517 \\
   &         &      & & & & 1.297 & 0.13520 \\
   &         &      & & & & 1.299 & 0.13523 \\
\hline
16 & 10.8910 & 300 & 1.23 & 0.13484 &  100 & 1.285 & 0.13482\\
     &               &        & 1.26 & 0.13487 & & 1.286 & 0.13484\\
     &               &        & 1.29 &               & & 1.287 & 0.13487\\
     &               &        & 1.32 &               & & 1.288 & 0.13489\\
\hline
\caption{\scriptsize Same caption as in tab.~\ref{tab:tuning_guess_np_new} but at
  scale P (see text).}
\label{tab:tuning_guess_p_new}
\end{longtable}
\newpage
\begin{longtable}[c]{|r|l||r|l|l|}
\hline
\multicolumn{5}{|c|}{Guess values for the tuning} \\
\multicolumn{5}{|c|}{2P scale} \\
\hline
\multicolumn{1}{|c|}{$L/a$} &
\multicolumn{1}{|c||}{$\beta$} &
\multicolumn{1}{|c|}{$N_{\textrm{conf}}$} &
\multicolumn{1}{|c|}{$z_{f}$ } &
\multicolumn{1}{|c|}{$\kappa$ } \\
\hline
16 & 12.0000 & 100 & 1.23 & 0.1335 \\
     &               &        & 1.24 & 0.1336 \\
     &               &        & 1.25 & 0.1337 \\
     &               &        & 1.26 & 0.1338 \\
\hline
\caption{\scriptsize Same caption as in tab.~\ref{tab:tuning_guess_np_new} but at
  scale 2P (see text). Here no separate tuning was performed for method (1).}
\label{tab:tuning_guess_2p_new}
\end{longtable}

\begin{longtable}[c]{|r|l||r|l|l|}
\hline
\multicolumn{5}{|c|}{Guess values for the tuning} \\
\multicolumn{5}{|c|}{PP scale} \\
\hline
\multicolumn{1}{|c|}{$L/a$} &
\multicolumn{1}{|c||}{$\beta$} &
\multicolumn{1}{|c|}{$N_{\textrm{conf}}$} &
\multicolumn{1}{|c|}{$z_{f}$ } &
\multicolumn{1}{|c|}{$\kappa$ } \\
\hline
 16 & 24.0000 & 80 & 1.11 & 0.1287 \\
      &               &      & 1.12 & 0.1288 \\
      &               &      & 1.13 & 0.1289 \\
      &               &      & 1.14 & 0.1290 \\
\hline
\caption{\scriptsize Same caption as in tab.~\ref{tab:tuning_guess_np_new} but at
  scale PP (see text). Here no separate tuning was performed for method (1).}
\label{tab:tuning_guess_pp_new}
\end{longtable}

\clearpage

\subsection{Tuning results}
\label{app:tuning_results}

\begin{table}[h]\tiny
\hspace{-1.0cm}
\begin{tabular}[l]{|l|l|l|l|l|l|l|l|l|l|}
\hline
\multicolumn{1}{|c|}{$\beta$} &
\multicolumn{1}{|c|}{$\kcr(1^{*})$ } &
\multicolumn{1}{|c|}{$\kcr(1)$ } &
\multicolumn{1}{|c|}{$\kcr(2)$ } &
\multicolumn{1}{|c|}{$\kcr(3)$ } &
\multicolumn{1}{|c|}{$\kcr(4)$ } &
\multicolumn{1}{|c|}{$\kcr(5)$ } &
\multicolumn{1}{|c|}{$\kcr(6)$ } &
\multicolumn{1}{|c|}{$\kcr(7)$ } &
\multicolumn{1}{|c|}{$\kcr(SF)$ } \\
\hline
\multicolumn{10}{|c|}{Hadronic scale: $L=1.436\, r_{0}$ ($\mu \sim 300\mathrm{ MeV}$)}\\
\hline
6.0219 & 0.153530\,(24) & 0.15353\,(66) & 0.15354\,(66) & 0.15352\,(67) & 0.15354\,(66) & 0.15354\,(66) & 0.15354\,(66) & 0.15353\,(66) & 0.153371\,(10)\\
6.1628 & 0.152134\,(17) & 0.15213\,(66) & 0.15214\,(66) & 0.15213\,(67) & 0.15214\,(66) & 0.15214\,(66) & 0.15214\,(66) &               & 0.152012\,(7) \\
6.2885 & 0.150815\,(22) & 0.15082\,(66) & 0.15082\,(66) & 0.15082\,(66) & 0.15082\,(65) & 0.15082\,(65) & 0.15082\,(65) & 0.15082\,(66) & 0.150752\,(10)\\
6.4956 & 0.148945\,(25) & 0.14894\,(34) & 0.14894\,(33) & 0.14893\,(34) & 0.14894\,(33) & 0.14894\,(33) & 0.14894\,(33) & 0.14894\,(33) & 0.148876\,(13)\\
6.6790 &                          & 0.14748\,(74) & 0.14748\,(74) & 0.14748\,(74) & 0.14748\,(73) & 0.14748\,(73) & 0.14748\,(73) & & \\
6.8187 &                          & 0.14645\,(41) & 0.14645\,(41) & 0.14645\,(42) & 0.14645\,(41) & 0.14645\,(41) & 0.14645\,(41) & 0.14645\,(41) & \\
\hline
\multicolumn{10}{|c|}{Intermediate scale: $\overline{g}^{2}=2.4484$ ($\mu \sim 1\mathrm{ GeV}$)} \\
\hline
7.0197 & 0.144501\,(13) & 0.14450\,(41) & 0.14450\,(41) & 0.14450\,(41) & 0.14450\,(41) & 0.14450\,(41) & 0.14450\,(41) & 0.14450\,(41) & 0.144454\,(7) \\
7.3551 & 0.143113\,(12) & 0.14311\,(29) & 0.14311\,(29) & 0.14311\,(29) & 0.14311\,(29) & 0.14311\,(29) & 0.14311\,(29) & 0.14311\,(29) & 0.143113\,(6) \\
7.6101 & 0.142112\,(13) & 0.14212\,(23) & 0.14212\,(23) & 0.14212\,(23) & 0.14212\,(23) & 0.14212\,(23) & 0.14212\,(23) & 0.14212\,(23) & 0.142107\,(6) \\
\hline
\multicolumn{10}{|c|}{Perturbative scale: $\overline{g}^{2}=0.9944$ ($\mu \sim 30\mathrm{ GeV}$)} \\
\hline
10.3000& 0.1354609\,(54) & & & & & & &                                                                                                         & 0.135457\,(5) \\
10.6086& 0.1351758\,(56) & & & & & & &                                                                                                         & 0.135160\,(4) \\
10.8910& 0.1348440\,(61) & 0.134844\,(93) & 0.134844\,(93) & 0.134844\,(93) & 0.134844\,(93) & 0.134844\,(93) & 0.134844\,(93) & 0.134844\,(93)& 0.134849\,(6) \\
\hline
\multicolumn{10}{|c|}{2P scale} \\
\hline
12.0000& & 0.13363\,(41) &  0.13363\,(41) &  0.13363\,(41) &  0.13363\,(41) &  0.13363\,(41) &  0.13363\,(41) &  0.13363\,(41) & \\
\hline
\multicolumn{10}{|c|}{PP scale} \\
\hline
24.0000& & 0.12877\,(15) & 0.12877\,(15) & 0.12877\,(15) & 0.12877\,(15) & 0.12877\,(15) & 0.12877\,(15) & 0.12877\,(15) & \\
\hline
\end{tabular}
\caption{\scriptsize Summary table of $\kcr$ for all beta values and
  tuning conditions, (1) to (7) (see sect.~\ref{ssec:Tuningconditions}
  for a description of all the methods). The data of column (1*)
  correspond to a separate analysis with method (1), using slightly
  different simulation parameters
  (cf. tab.~\ref{tab:tuning_guess_np_new},
  tab.~\ref{tab:tuning_guess_int_new} and tab.~\ref{tab:tuning_guess_p_new}).
  For reference, we also give $\kcr$
  for the SF~\cite{Guagnelli:2005zc,Guagnelli:2003hw,Guagnelli:2004za}.}
\label{tab:kc_beta}
\end{table}

\vspace{0.5cm}
\tiny
\begin{longtable}[b]{|l|l|l|l|l|l|l|l|l|}
\hline
\multicolumn{1}{|c|}{$\beta$} &
\multicolumn{1}{|c|}{$\zfcr(1^{*})$ } &
\multicolumn{1}{|c|}{$\zfcr(1)$ } &
\multicolumn{1}{|c|}{$\zfcr(2)$ } &
\multicolumn{1}{|c|}{$\zfcr(3)$ } &
\multicolumn{1}{|c|}{$\zfcr(4)$ } &
\multicolumn{1}{|c|}{$\zfcr(5)$ } &
\multicolumn{1}{|c|}{$\zfcr(6)$ } &
\multicolumn{1}{|c|}{$\zfcr(7)$ } \\
\hline
\multicolumn{9}{|c|}{Hadronic scale: $L=1.436\, r_{0}$}\\
\hline
6.0219 & 1.8090\,(32)& 1.8091\,(32)& 1.7946\,(34)& 1.8434\,(37)& 1.7656\,(27)& 1.7597\,(27)& 1.7835\,(40)& 1.7980\,(17)\\
6.1628 & 1.7920\,(30)& 1.7923\,(29)& 1.7820\,(31)& 1.8175\,(33)& 1.7541\,(25)& 1.7497\,(25)& 1.7687\,(38)& \\
6.2885 & 1.7664\,(51)& 1.7658\,(38)& 1.7573\,(40)& 1.7869\,(46)& 1.7312\,(34)& 1.7283\,(32)& 1.7408\,(56)& 1.7509\,(22)\\
6.4956 & 1.7212\,(83)& 1.7201\,(41)& 1.7132\,(44)& 1.7377\,(46)& 1.6929\,(35)& 1.6894\,(34)& 1.7053\,(63)& 1.7076\,(21)\\
6.6790 &                     & 1.6841\,(56)& 1.6789\,(59)& 1.6973\,(65)& 1.6582\,(52)& 1.6577\,(52)& 1.6600\,(90)& \\
6.8187 &                     & 1.6427\,(56)& 1.6381\,(60)& 1.6529\,(60)& 1.6253\,(51)& 1.6201\,(50)& 1.6421\,(88)& 1.6366\,(27)\\
\end{longtable}
\newpage
\scriptsize
\begin{longtable}[c]{|r|l|l|l|l|l|l|}
\hline
\multicolumn{7}{|c|}{$\Delta \zfcr(m)$ for different methods} \\
\multicolumn{7}{|c|}{Hadronic scale: $L=1.436\, r_{0}$} \\
\hline
\multicolumn{1}{|c|}{$L/a$} &
\multicolumn{1}{|c|}{2} &
\multicolumn{1}{|c|}{3} &
\multicolumn{1}{|c|}{4} &
\multicolumn{1}{|c|}{5} &
\multicolumn{1}{|c|}{6} &
\multicolumn{1}{|c|}{7} \\
\hline
  8 & 0.0145\,(47) & -0.0343\,(49) & 0.0435\,(42) & 0.0494\,(42) & 0.0256\,(51) & 0.0111\,(36) \\
10 & 0.0103\,(42) & -0.0252\,(44) & 0.0382\,(38) & 0.0426\,(38) & 0.0236\,(48) & \\
12 & 0.0085\,(55) & -0.0211\,(60) & 0.0346\,(51) & 0.0375\,(50) & 0.0250\,(68) & 0.0149\,(44) \\
16 & 0.0069\,(60) & -0.0176\,(62) & 0.0272\,(54) & 0.0307\,(53) & 0.0148\,(75) & 0.0125\,(46) \\
20 & 0.0052\,(81) & -0.0132\,(86) & 0.0259\,(76) & 0.0264\,(76) & 0.024\,(11)   & \\
24 & 0.0046\,(82) & -0.0102\,(82) & 0.0174\,(76) & 0.0226\,(75) & 0.001\,(10)   & 0.0061\,(62) \\
\hline
\caption{\scriptsize Differences of $\zfcr$, $\Delta \zfcr(m)$, determined from different
  methods. The differences are always $\zfcr(1)$ minus
  $\zfcr(m)$, obtained from any other method
  $m=2 \dots 7$. Scale NP.}
\label{tab:Delta_zf_1M_np_cl}
\end{longtable}

\begin{longtable}[c]{|c||r @{.} l|r @{.} l|}
\hline
\multicolumn{5}{|c|}{Continuum limit of $\Delta \zfcr(m)$.} \\
\multicolumn{5}{|c|}{Hadronic scale: $L=1.436\, r_{0}$} \\
\hline
\multicolumn{5}{|c|}{Fit: $\Delta \zfcr=b_{0}+b_{1}\frac{a}{L}$} \\
\hline
\multicolumn{1}{|c||}{m} &
\multicolumn{2}{|c|}{$b_{0}$} &
\multicolumn{2}{|c|}{$b_{1}$} \\
\hline
2 &   0&0006\,(99) &   0&10\,(12) \\
3 & -0&001\,(10)   & -0&24\,(13) \\
4 &   0&0076\,(91) &   0&31\,(11) \\
5 &   0&0096\,(91) &   0&33\,(11) \\
6 & -0&000\,(12)   &   0&25\,(15) \\
7 & -0&001\,(12)   &   0&20\,(18) \\
\hline
\caption{\scriptsize Continuum limit of the data presented in
  tab.~\ref{tab:Delta_zf_1M_np_cl}. We have performed a linear fit in
  $a/L$. The point $L/a=8$ is not included in the fit.}
\label{tab:Fit_Delta_zf_1M_np_cl}
\end{longtable}

\clearpage

\subsection{Scaling studies}
\label{app:scaling_studies}

\begin{longtable}[c]{|r|l|c|l|l|l|l|l|}
\hline
\multicolumn{8}{|c|}{${\mathcal R}_5^{1,2}$-even correlation functions} \\
\multicolumn{8}{|c|}{NP scale: $L=1.436\, r_{0}$ and $\boldsymbol{\theta}=(0,0,0)$} \\
\hline
\multicolumn{1}{|c|}{$L/a$} &
\multicolumn{1}{|c|}{$\beta$} &
\multicolumn{1}{|c|}{m} &
\multicolumn{1}{|c|}{$g_{\mathrm{P}_{-}}^{11}$ } &
\multicolumn{1}{|c|}{$g_{\mathrm{P}_{+}}^{11}$ } &
\multicolumn{1}{|c|}{$g_{\mathrm{V}_{0 -}}^{12}$ } &
\multicolumn{1}{|c|}{$g_{\mathrm{V}_{0 +}}^{12}$ } &
\multicolumn{1}{|c|}{$g_{1}^{11}$ } \\
\hline
\endhead
\hline
\caption[]{(continuing on the next page, caption below)}
\endfoot
\hline
\caption{Results for the ${\mathcal R}_5^{1,2}$-even correlation
  functions,
  $g_{\mathrm{P}_{\pm}}^{11}$,
  $g_{\mathrm{V}_{0 \pm}}^{12}$
  and
  $g_{1}^{11}$,
  at scale $L=1.436\, r_{0}$ and for $\boldsymbol{\theta}=(0,0,0)$.
  The data have been obtained via linear interpolations to the critical
  values of $\kappa$ and quadratic interpolations to the critical
  values of $z_{f}$ for all the tuning methods (1) to
  (7). See
  sect.~\ref{ssec:R512evenminus}
  and
  sect.~\ref{ssec:R512evenplus}
  for explanations. For the correlation functions $g_{\mathrm{P}_{+}}^{11}$ and $g_{\mathrm{V}_{0 +}}^{12}$ the first 
  error is statistical while the second, where available, contains the propagation of the error in $\zfcr$ (see text).}
\endlastfoot
  8 & 6.0219 & 1 & 3.637\,(14) & 0.02609\,(25)[82] & 1.9779\,(85) & 0.01279\,(16)[42] & 1.1857\,(53)\\
     &             & 2 & 3.643\,(14) & 0.02249\,(19)[63] & 1.9839\,(85) & 0.01090\,(12)[32] & 1.1883\,(52)\\
     &             & 3 & 3.619\,(14) & 0.03956\,(38)[164] & 1.9568\,(87) & 0.01990\,(24)[86] & 1.1767\,(54)\\
     &             & 4 & 3.652\,(13) & 0.01903\,(13)[17] & 1.9907\,(83) & 0.009143\,(79)[89] & 1.1915\,(51)\\
     &             & 5 & 3.653\,(13) & 0.01893\,(14)[14] & 1.9912\,(83) & 0.009112\,(80)[80] & 1.1918\,(51)\\
     &             & 6 & 3.647\,(13) & 0.02057\,(15)[53] & 1.9873\,(84) & 0.00991\,(10)[27] & 1.1899\,(52)\\
     &             & 7 & 3.642\,(14) & 0.02322\,(20) & 1.9827\,(85) & 0.01128\,(13) & 1.1878\,(52) \\
\hline
10 & 6.1628 & 1 & 3.605\,(13) & 0.02081\,(22)[93] & 1.8788\,(84) & 0.01028\,(13)[47] & 1.1136\,(51)\\
\end{longtable}

\begin{longtable}[c]{|r|l|c|l|l|l|l|l|}
\hline
\multicolumn{8}{|c|}{${\mathcal R}_5^{1,2}$-even correlation functions} \\
\multicolumn{8}{|c|}{I scale: $\overline{g}^{2}=2.4484$ and
  $\boldsymbol{\theta}=(0,0,0)$} \\
\hline
\multicolumn{1}{|c|}{$L/a$} &
\multicolumn{1}{|c|}{$\beta$} &
\multicolumn{1}{|c|}{m} &
\multicolumn{1}{|c|}{$g_{\mathrm{P}_{-}}^{11}$ } &
\multicolumn{1}{|c|}{$g_{\mathrm{P}_{+}}^{11}$ } &
\multicolumn{1}{|c|}{$g_{\mathrm{V}_{0 -}}^{12}$ } &
\multicolumn{1}{|c|}{$g_{\mathrm{V}_{0 +}}^{12}$ } &
\multicolumn{1}{|c|}{$g_{1}^{11}$ } \\
\hline
  8 & 7.0197 & 1 & 3.2342\,(83) & 0.010013\,(72)[84] & 2.2115\,(68) & 0.006152\,(52)[57] & 1.5371\,(48)\\
     &             & 2 & 3.2391\,(82) & 0.010043\,(76)[82] & 2.2170\,(68) & 0.006203\,(54)[62] & 1.5412\,(48)\\
     &             & 3 & 3.2240\,(85) & 0.011667\,(97)[237] & 2.1987\,(69) & 0.007214\,(71)[153] & 1.5286\,(49)\\
     &             & 4 & 3.2446\,(81) & 0.01110\,(10)[13]   & 2.2222\,(67) & 0.006959\,(69)[97] & 1.5456\,(47)\\
     &             & 5 & 3.2456\,(81) & 0.01148\,(11)[14]   & 2.2231\,(67) & 0.007222\,(74)[102] & 1.5464\,(47)\\
\end{longtable}

\clearpage

\begin{longtable}[c]{|r|l|c|l|l|l|l|l|}
\hline
\multicolumn{8}{|c|}{${\mathcal R}_5^{1,2}$-even correlation functions} \\
\multicolumn{8}{|c|}{P scale: $\overline{g}^{2}=0.9944$ and
  $\boldsymbol{\theta}=(0,0,0)$} \\
\hline
\multicolumn{1}{|c|}{$L/a$} &
\multicolumn{1}{|c|}{$\beta$} &
\multicolumn{1}{|c|}{m} &
\multicolumn{1}{|c|}{$g_{\mathrm{P}_{-}}^{11}$ } &
\multicolumn{1}{|c|}{$g_{\mathrm{P}_{+}}^{11}$ } &
\multicolumn{1}{|c|}{$g_{\mathrm{V}_{0 -}}^{12}$ } &
\multicolumn{1}{|c|}{$g_{\mathrm{V}_{0 +}}^{12}$ } &
\multicolumn{1}{|c|}{$g_{1}^{11}$ } \\
\hline
16 & 10.8910 & 1 & 3.0565\,(76) & 0.0006071\,(87)[89] & 2.4403\,(81) & 0.0004638\,(70)[72] & 1.9652\,(65)\\ 
     &               & 2 & 3.0568\,(76) & 0.0006251\,(90)[148] & 2.4407\,(81) & 0.0004782\,(72)[121] & 1.9656\,(65)\\ 
     &               & 3 & 3.0560\,(76) & 0.0006260\,(90)[164] & 2.4396\,(81) & 0.0004787\,(73)[129] & 1.9646\,(65)\\ 
     &               & 4 & 3.0574\,(75) & 0.000772\,(11)[32]   & 2.4414\,(81) & 0.0005952\,(88)[263] & 1.9663\,(65)\\ 
     &               & 5 & 3.0574\,(75) & 0.000782\,(11)[30]   & 2.4414\,(81) & 0.0006035\,(89)[245] & 1.9663\,(65)\\ 
     &               & 6 & 3.0573\,(75) & 0.000750\,(11)[49]   & 2.4414\,(81) & 0.0005775\,(86)[395] & 1.9662\,(65)\\ 
     &               & 7 & 3.0566\,(76) & 0.0006105\,(88) & 2.4405\,(81) & 0.0004665\,(70) & 1.9653\,(65)\\
\hline
\caption{Same caption as in tab.~\ref{tab:ecf_0.0_newtun_np}
  but at scale $\overline{g}^{2}=0.9944$.}
\label{tab:ecf_0.0_newtun_p}
\end{longtable}

\begin{longtable}[c]{|r|l|c|l|l|l|l|l|}
\hline
\multicolumn{8}{|c|}{${\mathcal R}_5^{1,2}$-even correlation functions} \\
\multicolumn{8}{|c|}{2P scale and $\boldsymbol{\theta}=(0,0,0)$} \\
\hline
\multicolumn{1}{|c|}{$L/a$} &
\multicolumn{1}{|c|}{$\beta$} &
\multicolumn{1}{|c|}{m} &
\multicolumn{1}{|c|}{$g_{\mathrm{P}_{-}}^{11}$ } &
\multicolumn{1}{|c|}{$g_{\mathrm{P}_{+}}^{11}$ } &
\multicolumn{1}{|c|}{$g_{\mathrm{V}_{0 -}}^{12}$ } &
\multicolumn{1}{|c|}{$g_{\mathrm{V}_{0 +}}^{12}$ } &
\multicolumn{1}{|c|}{$g_{1}^{11}$ } \\
\hline
16 & 12.0000 & 1 & 3.049\,(11) & 0.000537\,(16)[28] & 2.527\,(13) & 0.000429\,(13)[23] & 2.089\,(11)\\
     &               & 2 & 3.049\,(11) & 0.000569\,(17)[37] & 2.527\,(13) & 0.000455\,(14)[31] & 2.089\,(11)\\
     &               & 3 & 3.048\,(11) & 0.000508\,(15)[20] & 2.526\,(13) & 0.000404\,(12)[17] & 2.088\,(11)\\
     &               & 4 & 3.049\,(11) & 0.000754\,(21)[62] & 2.527\,(13) & 0.000609\,(18)[52] & 2.090\,(11)\\
     &               & 5 & 3.049\,(11) & 0.000726\,(21)[50] & 2.527\,(13) & 0.000586\,(17)[42] & 2.090\,(11)\\
     &               & 6 & 3.049\,(11) & 0.000813\,(23)[112] & 2.527\,(13) & 0.000658\,(19)[94] & 2.090\,(11)\\
     &               & 7 & 3.049\,(11) & 0.000535\,(16) & 2.526\,(13) & 0.000427\,(13) & 2.089\,(11)\\
\hline
\caption{Same caption as in tab.~\ref{tab:ecf_0.0_newtun_np}
  but at scale 2P.}
\label{tab:ecf_0.0_newtun_2p}
\end{longtable}

\begin{longtable}[c]{|r|l|c|l|l|l|l|l|}
\hline
\multicolumn{8}{|c|}{${\mathcal R}_5^{1,2}$-even correlation functions} \\
\multicolumn{8}{|c|}{PP scale and $\boldsymbol{\theta}=(0,0,0)$} \\
\hline
\multicolumn{1}{|c|}{$L/a$} &
\multicolumn{1}{|c|}{$\beta$} &
\multicolumn{1}{|c|}{m} &
\multicolumn{1}{|c|}{$g_{\mathrm{P}_{-}}^{11}$ } &
\multicolumn{1}{|c|}{$g_{\mathrm{P}_{+}}^{11}$ } &
\multicolumn{1}{|c|}{$g_{\mathrm{V}_{0 -}}^{12}$ } &
\multicolumn{1}{|c|}{$g_{\mathrm{V}_{0 +}}^{12}$ } &
\multicolumn{1}{|c|}{$g_{1}^{11}$ } \\
\hline
16 & 24.0000 & 1 & 3.0201\,(56) & 0.0002171\,(50)[61] & 2.7853\,(73) & 0.0001936\,(46)[57] & 2.5652\,(67)\\
     &               & 2 & 3.0201\,(56) & 0.0002289\,(52)[81] & 2.7854\,(73) & 0.0002045\,(48)[76] & 2.5653\,(67)\\
     &               & 3 & 3.0200\,(56) & 0.0002117\,(50)[50] & 2.7852\,(73) & 0.0001885\,(46)[46] & 2.5651\,(67)\\
     &               & 4 & 3.0202\,(56) & 0.0002566\,(56)[114] & 2.7855\,(73) & 0.0002301\,(51)[106] & 2.5655\,(67)\\
     &               & 5 & 3.0202\,(56) & 0.0002589\,(56)[99] & 2.7855\,(73) & 0.0002322\,(51)[92] & 2.5655\,(67)\\
     &               & 6 & 3.0202\,(56) & 0.0002528\,(55)[185] & 2.7855\,(73) & 0.0002266\,(50)[172] & 2.5654\,(67)\\
     &               & 7 & 3.0201\,(56) & 0.0002171\,(50) & 2.7853\,(73) & 0.0001935\,(46) & 2.5652\,(67)\\
\hline
\caption{Same caption as in tab.~\ref{tab:ecf_0.0_newtun_np}
  but at scale PP.}
\label{tab:ecf_0.0_newtun_pp}
\end{longtable}

\clearpage

\begin{longtable}[c]{|r|l|l|l|l|l|l|l|}
\hline
\multicolumn{8}{|c|}{${\mathcal R}_5^{1,2}$-even correlation functions} \\
\multicolumn{8}{|c|}{NP scale: $L=1.436\, r_{0}$ and
  $\boldsymbol{\theta}=(0.5,0.5,0.5)$} \\
\hline
\multicolumn{1}{|c|}{$L/a$} &
\multicolumn{1}{|c|}{$\beta$} &
\multicolumn{1}{|c|}{m} &
\multicolumn{1}{|c|}{$g_{\mathrm{P}_{-}}^{11}$ } &
\multicolumn{1}{|c|}{$g_{\mathrm{P}_{+}}^{11}$ } &
\multicolumn{1}{|c|}{$g_{\mathrm{V}_{0 -}}^{12}$ } &
\multicolumn{1}{|c|}{$g_{\mathrm{V}_{0 +}}^{12}$ } &
\multicolumn{1}{|c|}{$g_{1}^{11}$ } \\
\hline
\endhead
\hline
\caption[]{(continuing on the next page, caption below)}
\endfoot
\hline
\caption{Same caption as in tab.~\ref{tab:ecf_0.0_newtun_np}
  but at $\boldsymbol{\theta}=(0.5,0.5,0.5)$. The values denoted with (1*)
  have been obtained by direct simulations at the critical values of
  $\kappa$ and $z_{f}$. See
  sect.~\ref{sec:Tuning.TheStrategy} for explanation
  of (1*). For the correlation functions $g_{\mathrm{P}_{+}}^{11}$ and $g_{\mathrm{V}_{0 +}}^{12}$ the first 
  error is statistical while the second, where available, contains the propagation of the error in $\zfcr$ (see text).}
\endlastfoot
  8 & 6.0219 & 1* & 2.9412\,(86) & 0.04976\,(16) & 1.4129\,(64) & 0.02207\,(12) & 0.8461\,(39)\\
     &             & 1  & 2.937\,(10)   & 0.04978\,(21)[76] & 1.4098\,(64) & 0.02206\,(14)[34] & 0.8443\,(39)\\
     &             & 2  & 2.9421\,(98) & 0.04653\,(17)[61] & 1.4150\,(63) & 0.02060\,(12)[27] & 0.8466\,(39)\\
     &             & 3  & 2.920\,(10)   & 0.06126\,(30)[143] & 1.3929\,(65) & 0.02729\,(19)[66] & 0.8371\,(40)\\
     &             & 4  & 2.9510\,(95) & 0.04291\,(15)[24] & 1.4218\,(62) & 0.01900\,(10)[12] & 0.8498\,(38)\\
     &             & 5  & 2.9524\,(95) & 0.04264\,(16)[20] & 1.4226\,(62) & 0.01889\,(10)[11] & 0.8502\,(38)\\
     &             & 6  & 2.9459\,(97) & 0.04469\,(15)[56] & 1.4182\,(62) & 0.01978\,(11)[25] & 0.8480\,(39)\\
     &             & 7  & 2.9409\,(99) & 0.04721\,(18) & 1.4139\,(63) & 0.02090\,(13) & 0.8461\,(39)\\
\end{longtable}

\begin{longtable}[c]{|r|l|l|l|l|l|l|l|}
\hline
\multicolumn{8}{|c|}{${\mathcal R}_5^{1,2}$-even correlation functions} \\
\multicolumn{8}{|c|}{I scale: $\overline{g}^{2}=2.4484$ and
  $\boldsymbol{\theta}=(0.5,0.5,0.5)$} \\
\hline
\multicolumn{1}{|c|}{$L/a$} &
\multicolumn{1}{|c|}{$\beta$} &
\multicolumn{1}{|c|}{m} &
\multicolumn{1}{|c|}{$g_{\mathrm{P}_{-}}^{11}$ } &
\multicolumn{1}{|c|}{$g_{\mathrm{P}_{+}}^{11}$ } &
\multicolumn{1}{|c|}{$g_{\mathrm{V}_{0 -}}^{12}$ } &
\multicolumn{1}{|c|}{$g_{\mathrm{V}_{0 +}}^{12}$ } &
\multicolumn{1}{|c|}{$g_{1}^{11}$ } \\
\hline
 8 & 7.0197 & 1* & 2.5142\,(48) & 0.003341\,(79) & 1.4348\,(43) & 0.017953\,(66) & 1.0006\,(30)\\
     &             & 1  & 2.5143\,(55) & 0.003306\,(79)[119] & 1.4358\,(44) & 0.017784\,(65)[76] & 1.0013\,(31)\\
     &             & 2  & 2.5186\,(54) & 0.003279\,(86)[86] & 1.4399\,(44) & 0.017680\,(66)[66] & 1.0042\,(31)\\
     &             & 3  & 2.5056\,(56) & 0.003480\,(85)[247] & 1.4265\,(45) & 0.018668\,(71)[139] & 0.9953\,(31)\\
     &             & 4  & 2.5235\,(53) & 0.00332\,(10)[90]   & 1.4442\,(43) & 0.017964\,(72)[72] & 1.0075\,(31)\\
\end{longtable}

\clearpage
\begin{longtable}[c]{|r|l|l|l|l|l|l|l|}
\hline
\multicolumn{8}{|c|}{${\mathcal R}_5^{1,2}$-even correlation functions} \\
\multicolumn{8}{|c|}{P scale: $\overline{g}^{2}=0.9944$ and
  $\boldsymbol{\theta}=(0.5,0.5,0.5)$} \\
\hline
\multicolumn{1}{|c|}{$L/a$} &
\multicolumn{1}{|c|}{$\beta$} &
\multicolumn{1}{|c|}{m} &
\multicolumn{1}{|c|}{$g_{\mathrm{P}_{-}}^{11}$ } &
\multicolumn{1}{|c|}{$g_{\mathrm{P}_{+}}^{11}$ } &
\multicolumn{1}{|c|}{$g_{\mathrm{V}_{0 -}}^{12}$ } &
\multicolumn{1}{|c|}{$g_{\mathrm{V}_{0 +}}^{12}$ } &
\multicolumn{1}{|c|}{$g_{1}^{11}$ } \\
\hline
16 & 10.8910 & 1* & 2.2924\,(43) & 0.006800\,(15) & 1.4342\,(44) & 0.004191\,(14) & 1.1555\,(35)\\
     &               & 1  & 2.2921\,(44) & 0.006796\,(15)[16] & 1.4340\,(44) & 0.004181\,(14)[14] & 1.1553\,(35)\\
     &               & 2  & 2.2924\,(44) & 0.006796\,(16)[16] & 1.4343\,(44) & 0.004182\,(14)[14] & 1.1556\,(35)\\
     &               & 3  & 2.2916\,(45) & 0.006830\,(15)[25] & 1.4334\,(44) & 0.004202\,(14)[18] & 1.1549\,(35)\\
     &               & 4  & 2.2931\,(44) & 0.006872\,(17)[21] & 1.4350\,(44) & 0.004230\,(15)[18] & 1.1562\,(35)\\
     &               & 5  & 2.2931\,(44) & 0.006878\,(17)[21] & 1.4350\,(44) & 0.004234\,(15)[18] & 1.1563\,(35)\\
     &               & 6  & 2.2930\,(44) & 0.006859\,(17)[29] & 1.4349\,(44) & 0.004222\,(15)[22] & 1.1562\,(35)\\
     &               & 7  & 2.2922\,(44) & 0.006794\,(15) & 1.4341\,(44) & 0.004180\,(14) & 1.1554\,(35)\\
\hline
\caption{Same caption as in tab.~\ref{tab:ecf_0.5_newtun_np} but at
  scale $\overline{g}^{2}=0.9944$.}
\label{tab:ecf_0.5_newtun_p}
\end{longtable}

\begin{longtable}[c]{|r|l|c|l|l|l|l|l|}
\hline
\multicolumn{8}{|c|}{${\mathcal R}_5^{1,2}$-even correlation functions} \\
\multicolumn{8}{|c|}{2P scale and $\boldsymbol{\theta}=(0.5,0.5,0.5)$} \\
\hline
\multicolumn{1}{|c|}{$L/a$} &
\multicolumn{1}{|c|}{$\beta$} &
\multicolumn{1}{|c|}{m} &
\multicolumn{1}{|c|}{$g_{\mathrm{P}_{-}}^{11}$ } &
\multicolumn{1}{|c|}{$g_{\mathrm{P}_{+}}^{11}$ } &
\multicolumn{1}{|c|}{$g_{\mathrm{V}_{0 -}}^{12}$ } &
\multicolumn{1}{|c|}{$g_{\mathrm{V}_{0 +}}^{12}$ } &
\multicolumn{1}{|c|}{$g_{1}^{11}$ } \\
\hline
16 & 12.0000 & 1 & 2.2686\,(65) & 0.006722\,(22)[22] & 1.4486\,(62) & 0.004247\,(19)[20] & 1.1988\,(52)\\
     &               & 2 & 2.2688\,(65) & 0.006737\,(22)[25] & 1.4488\,(62) & 0.004257\,(19)[23] & 1.1990\,(52)\\
     &               & 3 & 2.2683\,(66) & 0.006712\,(21)[21] & 1.4483\,(62) & 0.004240\,(19)[19] & 1.1985\,(52)\\
     &               & 4 & 2.2694\,(65) & 0.006845\,(25)[40] & 1.4494\,(62) & 0.004327\,(21)[31] & 1.1996\,(51)\\
     &               & 5 & 2.2693\,(65) & 0.006828\,(25)[33] & 1.4493\,(62) & 0.004316\,(20)[27] & 1.1995\,(51)\\
     &               & 6 & 2.2695\,(65) & 0.006882\,(26)[71] & 1.4495\,(62) & 0.004351\,(21)[50] & 1.1997\,(51)\\
     &               & 7 & 2.2686\,(65) & 0.006721\,(22) & 1.4486\,(62) & 0.004246\,(19) & 1.1988\,(52)\\
\hline
\caption{Same caption as in tab.~\ref{tab:ecf_0.0_newtun_np} but at
  scale 2P and $\boldsymbol{\theta}=(0.5,0.5,0.5)$.}
\label{tab:ecf_0.5_newtun_2p}
\end{longtable}

\begin{longtable}[c]{|r|l|c|l|l|l|l|l|}
\hline
\multicolumn{8}{|c|}{${\mathcal R}_5^{1,2}$-even correlation functions} \\
\multicolumn{8}{|c|}{PP scale and $\boldsymbol{\theta}=(0.5,0.5,0.5)$} \\
\hline
\multicolumn{1}{|c|}{$L/a$} &
\multicolumn{1}{|c|}{$\beta$} &
\multicolumn{1}{|c|}{m} &
\multicolumn{1}{|c|}{$g_{\mathrm{P}_{-}}^{11}$ } &
\multicolumn{1}{|c|}{$g_{\mathrm{P}_{+}}^{11}$ } &
\multicolumn{1}{|c|}{$g_{\mathrm{V}_{0 -}}^{12}$ } &
\multicolumn{1}{|c|}{$g_{\mathrm{V}_{0 +}}^{12}$ } &
\multicolumn{1}{|c|}{$g_{1}^{11}$ } \\
\hline
16 & 24.0000 & 1 & 2.1966\,(29) & 0.006459\,(11)[11] & 1.5009\,(31) & 0.004391\,(10)[10] & 1.3851\,(28)\\
     &               & 2 & 2.1967\,(29) & 0.006459\,(11)[11] & 1.5011\,(31) & 0.004392\,(10)[11] & 1.3853\,(28)\\
     &               & 3 & 2.1964\,(29) & 0.006464\,(10)[11] & 1.5007\,(31) & 0.004395\,(11)[11] & 1.3850\,(28)\\
     &               & 4 & 2.1968\,(29) & 0.006468\,(11)[11] & 1.5013\,(31) & 0.004398\,(10)[11] & 1.3855\,(28)\\
     &               & 5 & 2.1969\,(29) & 0.006469\,(11)[11] & 1.5013\,(31) & 0.004399\,(10)[11] & 1.3855\,(28)\\
     &               & 6 & 2.1968\,(29) & 0.006467\,(11)[13] & 1.5012\,(31) & 0.004397\,(10)[12] & 1.3854\,(28)\\
     &               & 7 & 2.1966\,(29) & 0.006459\,(11) & 1.5009\,(31) & 0.004391\,(10) & 1.3851\,(28)\\
\hline
\caption{Same caption as in tab.~\ref{tab:ecf_0.0_newtun_np} but at
  scale PP and $\boldsymbol{\theta}=(0.5,0.5,0.5)$.}
\label{tab:ecf_0.5_newtun_pp}
\end{longtable}

\clearpage

\begin{longtable}[c]{|r|l|c|r @{.} l|r @{.} l|}
\hline
\multicolumn{7}{|c|}{${\mathcal R}_5^{1,2}$-odd correlation functions} \\
\multicolumn{7}{|c|}{NP scale: $L=1.436\, r_{0}$ and $\boldsymbol{\theta}=(0,0,0)$} \\
\hline
\multicolumn{1}{|c|}{$L/a$} &
\multicolumn{1}{|c|}{$\beta$} &
\multicolumn{1}{|c|}{m} &
\multicolumn{2}{|c|}{$g_{\mathrm{A}_{0 -}}^{11}$ } &
\multicolumn{2}{|c|}{$\overline{g}_{\mathrm{V}_{k -}}^{12}$ } \\
\endhead
\hline
\caption[]{(continuing on the next page, caption below)}
\endfoot
\hline
\caption{Results for the ${\mathcal R}_5^{1,2}$-odd correlation
 functions, $g_{\mathrm{A}_{0 -}}^{11}$
 and
 $\overline{g}_{\mathrm{V}_{k -}}^{12}$,
 at scale $L=1.436\, r_{0}$ and for $\boldsymbol{\theta}=(0,0,0)$.
 The data have been obtained via linear interpolations to the critical
 values of $\kappa$ and $\zf$ for all the tuning methods (1) to
 (7). The lines represent the cases where the corresponding
 quantity has been used as tuning condition.
 See sect.~\ref{ssec:R512odd}
 for explanations.}
\endlastfoot
 8 & 6.0219 & 1 & \multicolumn{2}{c|}{--------} & -0&171\,(10) \\
     &             & 2 &   0&0434\,(97) & -0&114\,(10) \\
     &             & 3 & -0&1031\,(95) & -0&306\,(10) \\
     &             & 4 &   0&1305\,(98) & \multicolumn{2}{c|}{--------}  \\
     &             & 5 &   0&1482\,(98) &   0&023\,(11) \\
     &             & 6 &   0&0767\,(97) & -0&070\,(11) \\
     &             & 7 &   0&0332\,(97) & -0&127\,(10)\\
\hline
10 & 6.1628 & 1 &  \multicolumn{2}{c|}{--------}  & -0&1449\,(93) \\
\end{longtable}
\newpage
\begin{longtable}[c]{|r|l|c|r|r|}
\hline
\multicolumn{5}{|c|}{${\mathcal R}_5^{1,2}$-odd correlation functions} \\
\multicolumn{5}{|c|}{I scale: $\overline{g}^{2}=2.4484$ and $\boldsymbol{\theta}=(0,0,0)$} \\
\hline
\multicolumn{1}{|c|}{$L/a$} &
\multicolumn{1}{|c|}{$\beta$} &
\multicolumn{1}{|c|}{m} &
\multicolumn{1}{|c|}{$g_{\mathrm{A}_{0 -}}^{11}$ } &
\multicolumn{1}{|c|}{$\overline{g}_{\mathrm{V}_{k -}}^{12}$ } \\
\hline
  8 & 7.0197 & 1 & \multicolumn{1}{c|}{--------} & -0.1134\,(64)\\
     &             & 2 &   0.0406\,(61) & -0.0640\,(65)\\
     &             & 3 & -0.0730\,(60) & -0.2015\,(62)\\
     &             & 4 &   0.0935\,(62) & \multicolumn{1}{c|}{--------}\\
     &             & 5 &   0.1048\,(63) &   0.0137\,(66)\\
     &             & 6 &   0.0659\,(62) & -0.0334\,(65)\\
     &             & 7 &   0.0044\,(61) & -0.1079\,(64)\\
\hline
12 & 7.3551 & 1 & \multicolumn{1}{c|}{--------} & -0.0845\,(72)\\
     &             & 2 &   0.0192\,(67) & -0.0610\,(72)\\
     &             & 3 & -0.0355\,(67) & -0.1277\,(71)\\
     &             & 4 &   0.0695\,(68) &   \multicolumn{1}{c|}{--------}\\
     &             & 5 &   0.0732\,(68) &   0.0048\,(73)\\
     &             & 6 &   0.0597\,(68) & -0.0118\,(72)\\
     &             & 7 &   0.0072\,(67) & -0.0757\,(72)\\
\hline
16 & 7.6101 & 1 &   \multicolumn{1}{c|}{--------} & -0.0645\,(80)\\
     &             & 2 &   0.0131\,(75) & -0.0487\,(80)\\
     &             & 3 & -0.0236\,(74) & -0.0933\,(80)\\
     &             & 4 &   0.0531\,(75) & \multicolumn{1}{c|}{--------}\\
     &             & 5 &   0.0572\,(75) &   0.0048\,(81)\\
     &             & 6 &   0.0431\,(75) & -0.0123\,(80)\\
     &             & 7 &   0.0101\,(75) & -0.0523\,(80)\\
\hline
\caption{Same caption as in tab.~\ref{tab:ocf_0.0_newtun_np} but at
  scale $\overline{g}^{2}=2.4484$.}
\label{tab:ocf_0.0_newtun_int}
\end{longtable}

\begin{longtable}[c]{|r|l|c|r|r|}
\hline
\multicolumn{5}{|c|}{${\mathcal R}_5^{1,2}$-odd correlation functions} \\
\multicolumn{5}{|c|}{P scale: $\overline{g}^{2}=0.9944$ and $\boldsymbol{\theta}=(0,0,0)$} \\
\hline
\multicolumn{1}{|c|}{$L/a$} &
\multicolumn{1}{|c|}{$\beta$} &
\multicolumn{1}{|c|}{m} &
\multicolumn{1}{|c|}{$g_{\mathrm{A}_{0 -}}^{11}$ } &
\multicolumn{1}{|c|}{$\overline{g}_{\mathrm{V}_{k -}}^{12}$ } \\
\hline
16 & 10.8910 & 1 & \multicolumn{1}{c|}{--------}  & -0.0363\,(43)\\
     &               & 2 &   0.0096\,(41) & -0.0255\,(43)\\
     &               & 3 & -0.0136\,(41) & -0.0516\,(43)\\
     &               & 4 &   0.0323\,(41) &   \multicolumn{1}{c|}{--------} \\
     &               & 5 &   0.0334\,(41) &   0.0012\,(44)\\
     &               & 6 &   0.0299\,(41) & -0.0027\,(44)\\
     &               & 7 &   0.0034\,(41) & -0.0325\,(43)\\
\hline
\caption{Same caption as in tab.~\ref{tab:ocf_0.0_newtun_np} but at
  scale $\overline{g}^{2}=0.9944$.}
\label{tab:ocf_0.0_newtun_p}
\end{longtable}
\newpage
\begin{longtable}[c]{|r|l|c|r|r|}
\hline
\multicolumn{5}{|c|}{${\mathcal R}_5^{1,2}$-odd correlation functions} \\
\multicolumn{5}{|c|}{2P scale and $\boldsymbol{\theta}=(0,0,0)$} \\
\hline
\multicolumn{1}{|c|}{$L/a$} &
\multicolumn{1}{|c|}{$\beta$} &
\multicolumn{1}{|c|}{m} &
\multicolumn{1}{|c|}{$g_{\mathrm{A}_{0 -}}^{11}$ } &
\multicolumn{1}{|c|}{$\overline{g}_{\mathrm{V}_{k -}}^{12}$ } \\
\hline
16 & 12.0000 & 1 &   \multicolumn{1}{c|}{--------}  & -0.0292\,(70)\\
     &               & 2 &   0.0059\,(67) & -0.0228\,(70)\\
     &               & 3 & -0.0081\,(67) & -0.0382\,(70)\\
     &               & 4 &   0.0266\,(67) &   \multicolumn{1}{c|}{--------} \\
     &               & 5 &   0.0242\,(67) & -0.0026\,(70)\\
     &               & 6 &   0.0315\,(67) &   0.0054\,(70)\\
     &               & 7 & -0.0004\,(67) & -0.0297\,(70)\\
\hline
\caption{Same caption as in tab.~\ref{tab:ocf_0.0_newtun_np} but at
  scale 2P.}
\label{tab:ocf_0.0_newtun_2p}
\end{longtable}

\begin{longtable}[c]{|r|l|c|r|r|}
\hline
\multicolumn{5}{|c|}{${\mathcal R}_5^{1,2}$-odd correlation functions} \\
\multicolumn{5}{|c|}{PP scale and $\boldsymbol{\theta}=(0,0,0)$} \\
\hline
\multicolumn{1}{|c|}{$L/a$} &
\multicolumn{1}{|c|}{$\beta$} &
\multicolumn{1}{|c|}{m} &
\multicolumn{1}{|c|}{$g_{\mathrm{A}_{0 -}}^{11}$ } &
\multicolumn{1}{|c|}{$\overline{g}_{\mathrm{V}_{k -}}^{12}$ } \\
\hline
16 & 24.0000 & 1 & \multicolumn{1}{c|}{--------}  & -0.0144\,(31)\\
     &               & 2 &   0.0057\,(31) & -0.0084\,(31)\\
     &               & 3 & -0.0066\,(31) & -0.0213\,(31)\\
     &               & 4 &   0.0137\,(31) & \multicolumn{1}{c|}{--------} \\
     &               & 5 &   0.0143\,(31) &   0.0005\,(31)\\
     &               & 6 &   0.0128\,(31) & -0.0010\,(31)\\
     &               & 7 & -0.0001\,(31) & -0.0144\,(31)\\
\hline
\caption{Same caption as in tab.~\ref{tab:ocf_0.0_newtun_np} but at
  scale PP.}
\label{tab:ocf_0.0_newtun_pp}
\end{longtable}

\begin{longtable}[c]{|r|l|l|r @{.} l|r @{.} l|}
\hline
\multicolumn{7}{|c|}{${\mathcal R}_5^{1,2}$-odd correlation functions} \\
\multicolumn{7}{|c|}{NP scale: $L=1.436\, r_{0}$ and $\boldsymbol{\theta}=(0.5,0.5,0.5)$} \\
\hline
\multicolumn{1}{|c|}{$L/a$} &
\multicolumn{1}{|c|}{$\beta$} &
\multicolumn{1}{|c|}{m} &
\multicolumn{2}{|c|}{$g_{\mathrm{A}_{0 -}}^{11}$ } &
\multicolumn{2}{|c|}{$\overline{g}_{\mathrm{V}_{k -}}^{12}$ } \\
\endhead
\hline
\caption[]{(continuing on the next page, caption below)}
\endfoot
\hline
\caption{Same caption as in tab.~\ref{tab:ocf_0.0_newtun_np}
  but at $\boldsymbol{\theta}=(0.5,0.5,0.5)$. The values denoted with (1*)
  have been obtained by direct simulations at the critical values of
  $\kappa$ and $z_{f}$. See
  sect.~\ref{sec:Tuning.TheStrategy} for explanation
  of (1*).}
\endlastfoot
  8 & 6.0219 & 1* & -0&0270\,(59)  & \multicolumn{2}{|c|}{}\\
     &             & 1  & -0&0307\,(71) & -0&1465\,(75) \\
     &             & 2  & \multicolumn{2}{c|}{--------}  & -0&1034\,(76) \\
     &             & 3  & -0&1031\,(70) & -0&2482\,(73) \\
     &             & 4  &   0&0612\,(73) & -0&0174\,(78) \\
     &             & 5  &   0&0736\,(73) &   \multicolumn{2}{c|}{--------}  \\
     &             & 6  &   0&0234\,(72) & -0&0705\,(77) \\
     &             & 7  & -0&0072\,(71) & -0&1135\,(76)\\
\end{longtable}
\newpage
\begin{longtable}[c]{|r|l|l|r|r|}
\hline
\multicolumn{5}{|c|}{${\mathcal R}_5^{1,2}$-odd correlation functions} \\
\multicolumn{5}{|c|}{I scale: $\overline{g}^{2}=2.4484$ and $\boldsymbol{\theta}=(0.5,0.5,0.5)$} \\
\hline
\multicolumn{1}{|c|}{$L/a$} &
\multicolumn{1}{|c|}{$\beta$} &
\multicolumn{1}{|c|}{m} &
\multicolumn{1}{|c|}{$g_{\mathrm{A}_{0 -}}^{11}$ } &
\multicolumn{1}{|c|}{$\overline{g}_{\mathrm{V}_{k -}}^{12}$ } \\
\hline
  8 & 7.0197 & 1* & -0.0378\,(31) & \\
     &             & 1  & -0.0263\,(41) & -0.0898\,(43) \\
     &             & 2  & \multicolumn{1}{c|}{--------}& -0.0550\,(43) \\
     &             & 3  & -0.0730\,(40) & -0.1521\,(41) \\
     &             & 4  &   0.0338\,(42) & -0.0098\,(44) \\
\end{longtable}
\newpage

\begin{longtable}[c]{|r|l|l|r|r|}
\hline
\multicolumn{5}{|c|}{${\mathcal R}_5^{1,2}$-odd correlation functions} \\
\multicolumn{5}{|c|}{P scale: $\overline{g}^{2}=0.9944$ and $\boldsymbol{\theta}=(0.5,0.5,0.5)$} \\
\hline
\multicolumn{1}{|c|}{$L/a$} &
\multicolumn{1}{|c|}{$\beta$} &
\multicolumn{1}{|c|}{m} &
\multicolumn{1}{|c|}{$g_{\mathrm{A}_{0 -}}^{11}$ } &
\multicolumn{1}{|c|}{$\overline{g}_{\mathrm{V}_{k -}}^{12}$ } \\
\hline
8   & 10.3000  & 1* & -0.0146\,(14) & \\
\hline
12 & 10.6086  & 1* & -0.0063\,(14) & \\
\hline
16 & 10.8910  & 1* & -0.0034\,(15) & \\
     &                & 1   & -0.0056\,(25) & -0.0256\,(27) \\
     &                & 2   & \multicolumn{1}{c|}{--------} & -0.0183\,(27) \\
     &                & 3   & -0.0136\,(25) & -0.0360\,(27) \\
     &                & 4   &   0.0134\,(25) & -0.0008\,(27) \\
     &                & 5   &   0.0140\,(25) & \multicolumn{1}{c|}{--------} \\
     &                & 6   &   0.0120\,(25) & -0.0027\,(27) \\
     &                & 7   & -0.0036\,(25) & -0.0230\,(27) \\
\hline
\caption{\scriptsize Same caption as in tab.~\ref{tab:ocf_0.5_newtun_np} but at
  scale $\overline{g}^{2}=0.9944$.}
\label{tab:ocf_0.5_newtun_p}
\end{longtable}

\begin{longtable}[c]{|r|l|c|r|r|}
\hline
\multicolumn{5}{|c|}{${\mathcal R}_5^{1,2}$-odd correlation functions} \\
\multicolumn{5}{|c|}{2P scale and $\boldsymbol{\theta}=(0.5,0.5,0.5)$} \\
\hline
\multicolumn{1}{|c|}{$L/a$} &
\multicolumn{1}{|c|}{$\beta$} &
\multicolumn{1}{|c|}{m} &
\multicolumn{1}{|c|}{$g_{\mathrm{A}_{0 -}}^{11}$ } &
\multicolumn{1}{|c|}{$\overline{g}_{\mathrm{V}_{k -}}^{12}$ } \\
\hline
16 & 12.0000 & 1 & -0.0034\,(40) & -0.0178\,(42) \\
     &               & 2 & \multicolumn{1}{c|}{--------} & -0.0135\,(42) \\
     &               & 3 & -0.0081\,(40) & -0.0239\,(42) \\
     &               & 4 &   0.0118\,(40) &   0.0019\,(42) \\
     &               & 5 &   0.0104\,(40) &   \multicolumn{1}{c|}{--------} \\
     &               & 6 &   0.0146\,(40) &   0.0055\,(42) \\
     &               & 7 & -0.0037\,(40) & -0.0181\,(42) \\
\hline
\caption{\scriptsize Same caption as in tab.~\ref{tab:ocf_0.0_newtun_np} but at
  scale 2P and $\boldsymbol{\theta}=(0.5,0.5,0.5)$.}
\label{tab:ocf_0.5_newtun_2p}
\end{longtable}

\begin{longtable}[c]{|r|l|c|r|r|}
\hline
\multicolumn{5}{|c|}{${\mathcal R}_5^{1,2}$-odd correlation functions} \\
\multicolumn{5}{|c|}{PP scale and $\boldsymbol{\theta}=(0.5,0.5,0.5)$} \\
\hline
\multicolumn{1}{|c|}{$L/a$} &
\multicolumn{1}{|c|}{$\beta$} &
\multicolumn{1}{|c|}{m} &
\multicolumn{1}{|c|}{$g_{\mathrm{A}_{0 -}}^{11}$ } &
\multicolumn{1}{|c|}{$\overline{g}_{\mathrm{V}_{k -}}^{12}$ } \\
\hline
16 & 24.0000 & 1 & -0.0031\,(17) & -0.0097\,(17) \\
     &               & 2 & \multicolumn{1}{c|}{--------} & -0.0058\,(17) \\
     &               & 3 & -0.0066\,(17) & -0.0141\,(17) \\
     &               & 4 &   0.0043\,(17) & -0.0004\,(17) \\
     &               & 5 &   0.0046\,(17) & \multicolumn{1}{c|}{--------} \\
     &               & 6 &   0.0038\,(17) & -0.0010\,(17) \\
     &               & 7 & -0.0031\,(17) & -0.0097\,(17) \\
\hline
\caption{\scriptsize Same caption as in tab.~\ref{tab:ocf_0.0_newtun_np} but at
  scale PP and $\boldsymbol{\theta}=(0.5,0.5,0.5)$.}
\label{tab:ocf_0.5_newtun_pp}
\end{longtable}

\begin{longtable}[c]{|c||r @{.} l|r @{.} l||r @{.} l|r @{.} l|}
\hline
\multicolumn{9}{|c|}{Continuum limit of $g_{\mathrm{A}_{0 -}}^{11}$. NP scale} \\
\hline
\multicolumn{1}{|c||}{}& \multicolumn{4}{|c||}{$\boldsymbol{\theta}=(0,0,0)$} & \multicolumn{4}{|c|}{$\boldsymbol{\theta}=(0.5,0.5,0.5)$} \\
\hline
\multicolumn{1}{|c||}{m} &
\multicolumn{2}{|c|}{$b_{0}$ } &
\multicolumn{2}{|c||}{$b_{1}$ } &
\multicolumn{2}{|c|}{$b_{0}$ } &
\multicolumn{2}{|c|}{$b_{1}$ } \\
\hline
1 & \multicolumn{2}{c|}{--------} &   \multicolumn{2}{c||}{--------} &   0&001\,(14) & -0&22\,(17)\\
2 & -0&001\,(18) &   0&31\,(23) & \multicolumn{2}{c|}{--------}  &   \multicolumn{2}{c|}{--------} \\
3 &   0&003\,(18) & -0&75\,(23) &   0&004\,(14) & -0&76\,(17)\\
4 &   0&008\,(18) &   1&02\,(23) &   0&007\,(14) &   0&51\,(17)\\
5 &   0&013\,(18) &   1&09\,(23) &   0&010\,(14) &   0&56\,(17)\\
6 & -0&006\,(18) &   0&79\,(23) & -0&003\,(14) &   0&34\,(17)\\
7 & -0&005\,(29) &   0&57\,(42) & -0&005\,(21) &   0&22\,(31)\\
\hline
\caption{\scriptsize Continuum limit of $g_{\mathrm{A}_{0 -}}^{11}$ for all
  definitions of $\zfcr$ from (1) to (7). Scale NP.
  Results are shown for
  $\boldsymbol{\theta}=(0,0,0)$ and $\boldsymbol{\theta}=(0.5,0.5,0.5)$,
  as obtained from linear fits of the data in
  tab.~\ref{tab:ocf_0.0_newtun_np} and 
  tab.~\ref{tab:ocf_0.5_newtun_np}, respectively.
  The fits are linear in $a/L$ and
  the point $L/a=8$ is excluded in all cases.
  The lines represent the cases where the corresponding
  quantity has been used as tuning condition.
  See
  sect.~\ref{ssec:R512odd}
  for more explanations.}
\label{tab:ga0_cl_np_fit}
\end{longtable}

\begin{longtable}[c]{|c||r @{.} l|r @{.} l||r @{.} l|r @{.} l|}
\hline
\multicolumn{9}{|c|}{Continuum limit of $\overline{g}_{\mathrm{V}_{k -}}^{12}$. NP scale} \\
\hline
\multicolumn{1}{|c||}{}& \multicolumn{4}{|c||}{$\boldsymbol{\theta}=(0,0,0)$} & \multicolumn{4}{|c|}{$\boldsymbol{\theta}=(0.5,0.5,0.5)$} \\
\hline
\multicolumn{1}{|c||}{m} &
\multicolumn{2}{|c|}{$b_{0}$ } &
\multicolumn{2}{|c||}{$b_{1}$ } &
\multicolumn{2}{|c|}{$b_{0}$ } &
\multicolumn{2}{|c|}{$b_{1}$ } \\
\hline
1 & -0&008\,(21) & -1&37\,(26) & -0&012\,(16) & -1&12\,(20)\\
2 & -0&011\,(21) & -0&96\,(26) & -0&013\,(16) & -0&81\,(20)\\
3 & -0&002\,(20) & -2&39\,(26) & -0&008\,(16) & -1&89\,(19)\\
4 & \multicolumn{2}{c|}{--------} &  \multicolumn{2}{c||}{--------} & -0&004\,(16) & -0&07\,(20)\\
5 &   0&005\,(21) &   0&10\,(26) &   \multicolumn{2}{c|}{--------} & \multicolumn{2}{c|}{--------}\\
6 & -0&018\,(21) & -0&32\,(26) & -0&020\,(16) & -0&31\,(20)\\
7 & -0&002\,(33) & -0&81\,(48) & -0&019\,(25) & -0&51\,(36)\\
\hline
\caption{\scriptsize Same caption as in tab.~\ref{tab:ga0_cl_np_fit} but for
  $\overline{g}_{\mathrm{V}_{k -}}^{12}$.}
\label{tab:gak_cl_np_fit}
\end{longtable}
\clearpage

\newpage

\end{appendix}
\newpage
\bibliographystyle{h-elsevier}    
\bibliography{paper1}      
\end{document}